\documentclass[lettersize,journal]{IEEEtran}
\usepackage{amsmath,amssymb,amsfonts}
\usepackage{algorithmic}
\usepackage{array}
\usepackage[caption=false,font=normalsize,labelfont=sf,textfont=sf]{subfig}
\usepackage{textcomp}
\usepackage{stfloats}
\usepackage{url}
\usepackage{verbatim}
\usepackage{graphicx}
\def\BibTeX{{\rm B\kern-.05em{\sc i\kern-.025em b}\kern-.08em
		T\kern-.1667em\lower.7ex\hbox{E}\kern-.125emX}}
\usepackage{balance}

\usepackage{cite}
\usepackage{xcolor}

\usepackage{subfloat}
\usepackage{epstopdf}

\newtheorem{lemma}{Lemma}
\newtheorem{proposition}{Proposition}
\newtheorem{remark}{Remark}

\begin{document}
	\include{header}
	
	\title{ISAC Meets SWIPT: Multi-functional Wireless Systems Integrating Sensing, Communication, and Powering}
	
	\author{Yilong Chen, Haocheng Hua, Jie Xu, and Derrick Wing Kwan Ng \\
		\thanks{Part of this paper has been presented at the IEEE International Conference on Communications (ICC), Rome, Italy, 28 May - 1 June, 2023 \cite{chen2022transmit}.}
		\thanks{Y. Chen, H. Hua, and J. Xu are with the School of Science and Engineering (SSE) and the Future Network of Intelligence Institute (FNii), The Chinese University of Hong Kong (Shenzhen), Shenzhen, China (e-mail: yilongchen@link.cuhk.edu.cn, haochenghua@link.cuhk.edu.cn, xujie@cuhk.edu.cn).} 
		\thanks{D. W. K. Ng is with School of Electrical Engineering and Telecommunications, the University of New South Wales, Sydney, Australia (e-mail: w.k.ng@unsw.edu.au).}
		\thanks{J. Xu is the corresponding author.}
	}

	
	\maketitle
	
	\begin{abstract}
		This paper unifies integrated sensing and communication (ISAC) and simultaneous wireless information and power transfer (SWIPT), by investigating a new multi-functional multiple-input multiple-output (MIMO) system that integrates wireless sensing, communication, and powering. In this system, a multi-antenna hybrid access point (H-AP) transmits wireless signals to communicate with a multi-antenna information decoding (ID) receiver, wirelessly charges a multi-antenna energy harvesting (EH) receiver, and performs radar target sensing based on the echo signal concurrently. Under this setup, we aim to reveal the fundamental performance tradeoff limits among sensing, communication, and powering, in terms of the estimation Cram{\'e}r-Rao bound (CRB), achievable communication rate, and harvested energy, respectively. In particular, we consider two different target models for radar sensing, namely the point and extended targets, for which we are interested in estimating the target angle and the complete target response matrix, respectively. For both models, we define the achievable CRB-rate-energy (C-R-E) region and characterize its Pareto boundary by maximizing the achievable rate at the ID receiver, subject to the estimation CRB requirement for target sensing, the minimum harvested energy requirement at the EH receiver, and the maximum transmit power constraint at the H-AP. We obtain partitionable optimal transmit covariance matrix solutions to the two formulated problems by applying advanced convex optimization techniques. The numerical results demonstrate the optimal C-R-E region boundary achieved by our proposed design, as compared to the benchmark schemes based on time division and eigenmode transmission (EMT).
	\end{abstract}


\section{Introduction}

Future sixth-generation (6G) wireless networks are expected to support various new intelligent Internet-of-Things (IoT) applications such as smart homes, smart logistics, industrial automation, and smart healthcare \cite{cui2021integrating, saad2020vision, tong20216g}. Towards this end, wireless networks need to incorporate billions of low-power IoT devices and support their localization, sensing, communication, computation, and control in a sustainable manner \cite{saad2020vision}. According to the projection in \cite{tong20216g}, 6G networks are required to provide peak and user-experienced data rates of 1 Tbps and 10-100 Gbps, respectively, localization accuracy of 1 cm indoors and 50 cm outdoors, and battery lifetime of up to 20 years for low-power IoT devices. As a result, innovative wireless techniques are in high demand to satisfy these stringent key performance indicators (KPIs). 

On the one hand, by integrating the radio-frequency (RF)-based wireless power transfer (WPT) \cite{zeng2017communications} and wireless communications, simultaneous wireless information and power transfer (SWIPT) has recently emerged as an efficient solution for providing ubiquitous energy supplies to massive IoT devices, enabling their battery-free operation \cite{clerckx2019fundamentals}. In SWIPT, the same wireless signals are reused for simultaneously delivering information and energy to information decoding (ID) and energy harvesting (EH) receivers, respectively. On the other hand, integrated sensing and communication (ISAC) has been recognized as an enabling technique for 6G networks to provide both sensing and communication functionalities \cite{liu2022integrated}, in which the same wireless signals can be exploited for not only delivering information, but also sensing surrounding targets and environments based on the echo signals \cite{cui2021integrating}. With the recent advancements in these areas, we envision that future 6G networks will integrate both SWIPT and ISAC, evolving towards new multi-functional wireless systems, which can provide sensing, communication, and powering capabilities at the same time. 
These multi-functional wireless systems are expected to significantly enhance the utilization efficiency of scarce spectrum resources and densely deployed base station (BS) infrastructures, aiming for supporting emerging IoT applications, such as smart city and industrial automation. In these applications, millions of IoT or sensor devices are deployed to monitor the environment. In particular, these devices require sustainable power supplies to send their sensed data to the BS via uplink, while the BS needs to provide control information for these devices via downlink and localize them for advanced processing \cite{cui2021integrating}. In such scenarios, unifying ISAC and SWIPT can efficiently facilitate the localization and data collection of massive low-power devices, thereby reducing the network complexity. On the one hand, by adopting WPT, IoT nodes can utilize the harvested wireless power to support their sensing and communication, thus prolonging the lifetime of ISAC networks. On the other hand, sensing environmental targets or IoT devices can help to achieve adaptive resource allocation for SWIPT systems.

In the existing literature, there have been extensive prior works investigating the transmit optimization for SWIPT (e.g., \cite{zhang2013mimo, xu2014multiuser, luo2015capacity, park2013joint, ng2014robust, boshkovska2015practical}) and ISAC (e.g., \cite{xiong2022flowing, hua2022mimo, liu2018mumimo, liu2020joint, hua2021optimal, zhang2019multibeam, xu2021rate, yin2022rate, wang2022noma}) independently. For instance, the authors in \cite{zhang2013mimo} first studied a multiple-input multiple-output (MIMO) SWIPT system with an ID receiver and an EH receiver, in which the transmit covariance matrix at the BS was designed to optimally balance the tradeoff between the communication rate at the ID receiver versus the harvested energy at the EH receiver. This design was then extended to the broadcast channel with multiple ID receivers and multiple EH receivers, by considering the low-complexity linear transmit beamforming \cite{xu2014multiuser} and the capacity-achieving dirty paper coding \cite{luo2015capacity}, respectively. Furthermore, other prior works investigated SWIPT under different setups, e.g., in interference channels \cite{park2013joint} and secrecy communications \cite{ng2014robust}, or in the case with non-linear EH receivers \cite{boshkovska2015practical}. On the other hand, for ISAC systems, the works in \cite{xiong2022flowing} and \cite{hua2022mimo} considered the basic setup with a multi-antenna BS, a multi-antenna ID receiver, and a sensing target, in which the transmit strategies at the BS were optimized to balance the communication rate versus the estimation Cram{\'e}r-Rao bound (CRB) as the performance metric for target sensing. Furthermore, the authors in \cite{liu2018mumimo, liu2020joint, hua2021optimal} studied the ISAC system with multiple communication users and sensing targets, in which the transmit beamforming design at the BS was optimized to balance the communication and sensing performances, in terms of the signal-to-interference-plus-noise ratio (SINR) and radar beampatterns, respectively. In addition, the work \cite{zhang2019multibeam} developed a multi-beam transmission framework for ISAC exploiting analog antenna arrays. Furthermore, advanced multiple access techniques such as rate-splitting multiple access (RSMA) \cite{xu2021rate, yin2022rate} and non-orthogonal multiple access (NOMA) \cite{wang2022noma} were adopted to enhance ISAC performance.

Different from the prior works investigating ISAC and SWIPT independently, this paper studies a multi-functional MIMO system unifying ISAC and SWIPT, in which a multi-antenna hybrid access point (H-AP) transmits wireless signals to simultaneously deliver information to a multi-antenna ID receiver, provide energy supply to a multi-antenna EH receiver, and estimate a sensing target based on the echo signals. For such a multi-functional wireless system,\footnote{Practical IoT networks consist of millions of IoT devices, each having one or more of the three functionalities. As an initial attempt, this paper mainly considers the separate CRB-rate-energy (C-R-E) setup with an ID receiver, an EH receiver, and a target. We also investigate the co-located C-R/C-E (the ID/EH receiver also serves as a target) and R-E (a receiver performs both ID and EH functionalities) setups as special cases. The interesting scenario with multiple ID/EH receivers and targets is left for future work.} we aim to optimize the fundamental performance tradeoffs among sensing, communication, and powering, in terms of the estimation CRB, the achievable communication rate, and the harvested energy, respectively.
To address this, how to design the transmit strategies at the multi-antenna H-AP is essential. This problem, however, is particularly challenging, due to the following reasons. First, MIMO radar sensing, communication, and WPT are generally designed based on distinct objectives and follow different principles. In particular, for MIMO radar sensing, it is known that isotropic transmission, exploiting an identity sample covariance matrix, is beneficial for leveraging waveform diversity to enhance sensing performance \cite{li2008range, bekkerman2006target}. By contrast, for point-to-point MIMO communication with perfect channel state information (CSI) at transceivers, it is well established in \cite{telatar1999capacity} that eigenmode transmission (EMT), which involves implementing singular value decomposition (SVD) on the MIMO communication channel, along with water-filling (WF) power allocation over decomposed parallel subchannels, is optimal for maximizing MIMO channel capacity. As for MIMO WPT with linear radio frequency (RF)-to-direct current (DC) conversion efficiency, it is shown in \cite{zhang2013mimo} that the strongest EMT based on the MIMO WPT channel is optimal for maximizing harvested energy. Next, MIMO radar sensing, communication, and WPT systems handle inter-signal-stream interference in different manners. Specifically, in MIMO radar sensing, transmitting different independent probing signal beams is beneficial in reflecting target information from different perspectives. As in MIMO communication, inter-signal-stream interference is detrimental, which needs to be mitigated via proper signal processing techniques. By contrast, in MIMO WPT, inter-signal-stream interference is beneficial, as it can be harvested by RF-band EH receivers. Due to these differences in sensing, communication, and powering, finding the optimal transmission strategy for multi-functional wireless MIMO systems to balance their performance tradeoffs becomes an important but challenging problem. These motivate our study in this work.

In this paper, we consider the multi-functional MIMO system by focusing on two different target models for radar sensing, namely the point and extended targets, respectively. For the point target case, the H-AP aims to estimate the target reflection coefficient and the target angle as unknown parameters, while the target angle estimation CRB is considered as the sensing performance metric. As for the extended target case, the H-AP aims to estimate the complete target response matrix and the corresponding matrix estimation CRB is considered. The main results of this paper are summarized as follows.
\begin{itemize}
	\item We characterize the complete C-R-E regions of the multi-functional MIMO system for the point and extended target cases, which are defined as the set of all C-R-E pairs that are simultaneously achievable by sensing, communication, and WPT. Towards this end, we first identify three vertices on the Pareto boundary surface of each C-R-E region corresponding to the CRB minimization (C-min), rate maximization (R-max), and energy maximization (E-max), respectively, as well as three edges characterizing the optimal C-R, R-E, and C-E tradeoffs, respectively. Next, for acquiring the remaining points on the two Pareto boundary surfaces, we formulate two novel MIMO rate maximization problems, each subject to the corresponding estimation CRB requirement for target sensing, the harvested energy requirement at the EH receiver, and the maximum transmit power constraint at the H-AP.

	\item We derive partitionable optimal solutions to the two CRB-and-energy-constrained MIMO rate maximization problems using advanced convex optimization techniques. It is shown that for the point target case, the optimal transmit covariance matrix solution follows the EMT structure based on a composite channel matrix, combined with a WF-like power allocation. Specifically, this solution is generally divided into two parts: one serving the triple roles of communication, sensing, and powering, and the other dedicated to sensing and powering only. By contrast, for the extended target case, the optimal transmit covariance matrix is obtained by applying SVD to the combined ID and EH channel matrix. In particular, this solution is full rank and consists of two parts arranged in a block diagonal form: one for balancing the tradeoff among the three functionalities, and the other solely dedicated to sensing. Furthermore, we investigate the performance tradeoff of the three functionalities for special cases with co-located C-R, C-E, and R-E, respectively.

	\item We provide numerical results to validate the C-R-E region boundaries achieved by the optimal solutions for the multi-functional MIMO system. For the point target case, the resultant C-R-E tradeoff is shown to highly depend on the correlations among ID, EH, and sensing channels. As for the extended target case, the resultant C-R-E tradeoff is found to be influenced by system configuration such as the number of antennas at the H-AP. Furthermore, it is shown that our proposed optimal designs significantly outperform the benchmark schemes based on time division, EMT over the ID channel, and EMT over the combined ID and EH channels.
\end{itemize}

The remainder of this paper is organized as follows. Section II presents the investigated multi-functional wireless MIMO system model by considering the point and extended target cases, respectively. Section III characterizes the C-R-E regions of the multi-functional MIMO system for the two cases, by first identifying three vertices and three edges of the Pareto boundary surfaces, followed by the formulation of two CRB-and-energy-constrained MIMO rate maximization problems to determine all remaining boundary surface points. Sections IV and V present the optimal solutions to the two formulated problems for the point and extended target cases, respectively. Section VI provides numerical results to validate the performance of our proposed designs. Finally, Section VII concludes this paper.

\textit{Notations:} Boldface letters are used for vectors (lower-case) and matrices (upper-case). For a square matrix \(\boldsymbol{A}\), \(\mathrm{tr}(\boldsymbol{A})\), \(\det(\boldsymbol{A})\), and \(\boldsymbol{A}^{-1}\) denote its trace, determinant, and inverse, respectively, while \(\boldsymbol{A} \succeq \boldsymbol{0}\) and \(\boldsymbol{A} \succ \boldsymbol{0}\) mean that \(\boldsymbol{A}\) is positive semi-definite and positive definite, respectively. For an arbitrary-size matrix \(\boldsymbol{A}\), \(\mathrm{rank}(\boldsymbol{A})\), \(\boldsymbol{A}^\dag\), \(\boldsymbol{A}^T\), \(\boldsymbol{A}^H\), and \(\mathcal{R}(\boldsymbol{A})\) denote its rank, conjugate, transpose, conjugate transpose, and range space, respectively. For a vector \(\boldsymbol{a}\), \(\|\boldsymbol{a}\|\) denotes its Euclidean norm. For a complex number \(a\), \(|a|\) denotes its magnitude. For a real number \(a\), \((a)^+ \triangleq \max(a,0)\). \(\mathrm{diag}(a_1, \dots, a_N)\) denotes a diagonal matrix whose diagonal entries are \(a_1, \dots, a_N\). \(\boldsymbol{I}_N\) denotes the identity matrix with dimension \(N \times N\). \(\mathbb{C}^{M \times N}\), \(\mathbb{S}^{N}\), and \(\mathbb{S}_+^{N}\) denote the spaces of \(M \times N\) complex matrices, \(N \times N\) Hermitian matrices, and \(N \times N\) complex positive semidefinite matrices, respectively. \(\mathrm{span}\{\cdot\}\) denotes the linear span. \(\mathbb{E}[\cdot]\) denotes the statistic expectation. $\otimes$ denotes the Kronecker product. \(j = \sqrt{-1}\) is the imaginary unit. 

\section{System Model}

\begin{figure}[tb]
	\centering
	{\includegraphics[width=0.30\textwidth]{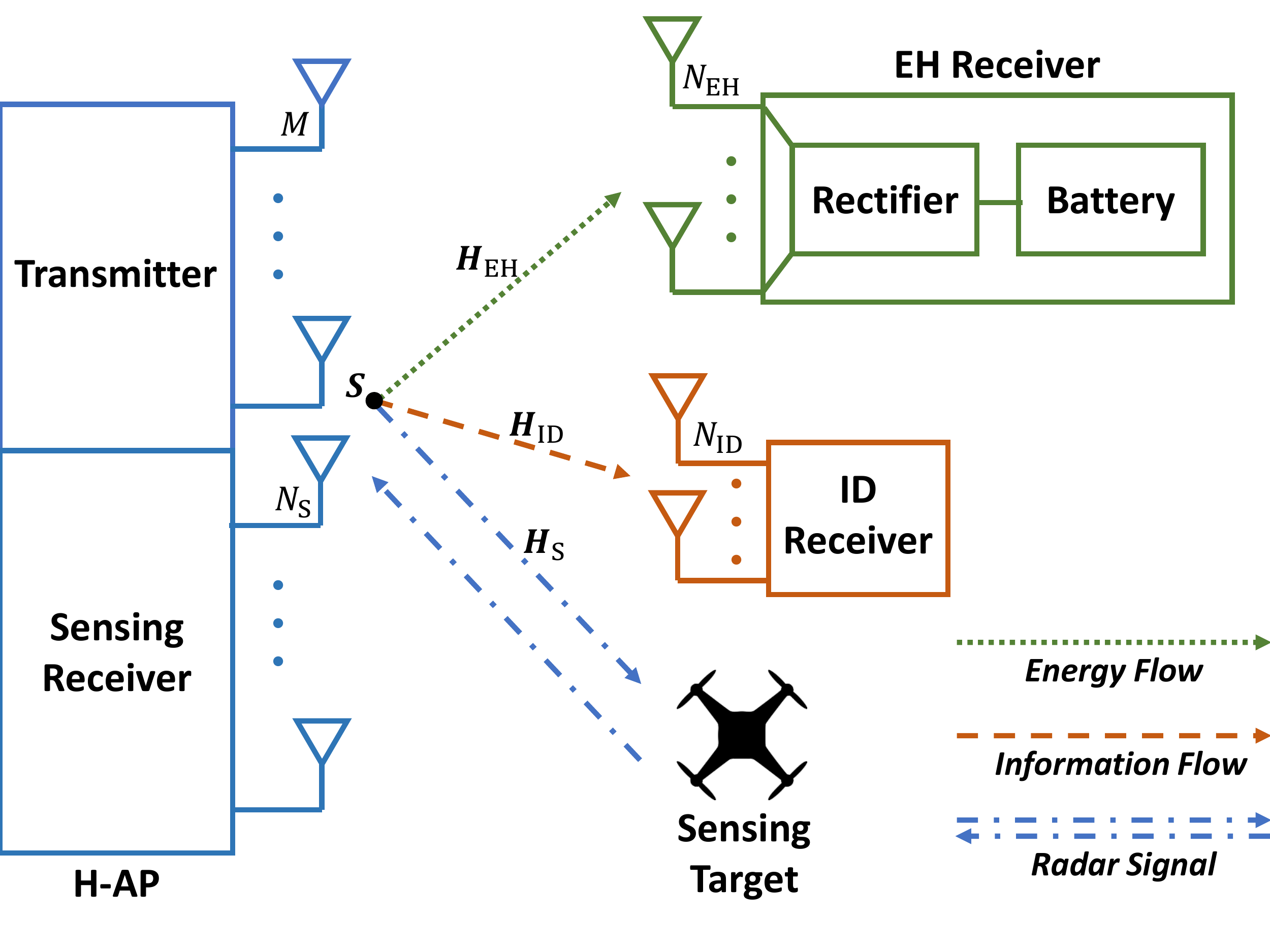}}
	\caption{The proposed multi-functional wireless system unifying ISAC and SWIPT.}
\end{figure}
This paper considers a multi-functional MIMO system. Fig. 1 shows the separate C-R-E setup, which consists of an H-AP, an EH receiver, an ID receiver, and a sensing target to be estimated. The H-AP is equipped with a uniform linear array (ULA) of \(M > 1\) transmit antennas and \(N_{\text{S}} > M\) receive antennas for sensing.\footnote{For radar estimation tasks, the number of receive antennas should be greater than that of transmit antennas to prevent any potential information loss of the sensed targets \cite{liu2021cramer}.} The EH and ID receivers are equipped with \(N_{\text{EH}} \ge 1\) and \(N_{\text{ID}} \ge 1\) receive antennas, respectively.

We consider a quasi-static narrowband channel model, in which the wireless channels remain invariant over the interested transmission block consisting of \(L\) symbols, where $L$ is assumed to be sufficiently large. Let \(\boldsymbol{H}_{\text{EH}} \in \mathbb{C}^{N_{\text{EH}} \times M}\) and \(\boldsymbol{H}_{\text{ID}} \in \mathbb{C}^{N_{\text{ID}} \times M}\) denote the channel matrices from the H-AP to the EH receiver and the ID receiver, respectively. Let \(\boldsymbol{H}_{\text{S}} \in \mathbb{C}^{N_{\text{S}} \times M}\) denote the target response matrix from the H-AP transmitter to the sensing target to the H-AP receiver, which will be specified later for the point and extended target models, respectively. To reveal the fundamental performance upper bound limits, it is assumed that the H-AP perfectly knows the CSI of \(\boldsymbol{H}_{\text{EH}}\) and \(\boldsymbol{H}_{\text{ID}}\), and the ID receiver perfectly knows \(\boldsymbol{H_{\text{ID}}}\) to facilitate the transmit optimization, as commonly assumed in the literature, e.g., \cite{zhang2013mimo, xu2014multiuser, luo2015capacity, park2013joint, ng2014robust, boshkovska2015practical}.

At each symbol \(l \in \{1, \dots, L\}\), let \(\boldsymbol{x}(l) \in \mathbb{C}^{M \times 1}\) denote the transmitted signal at the H-AP. The transmitted signal over the whole block is expressed as
\begin{equation}
	\boldsymbol{X} = \big[\boldsymbol{x}(1), \dots, \boldsymbol{x}(L)\big] \in \mathbb{C}^{M \times L}.
\end{equation}
Without loss of optimality, we consider capacity-achieving Gaussian signaling such that \(\boldsymbol{x}(l), \forall l\), are assumed to follow circularly symmetric complex Gaussian (CSCG) random vectors with zero mean and covariance matrix \(\boldsymbol{S} = \mathbb{E}\big[\boldsymbol{x}(l) \boldsymbol{x}(l)^H\big] \succeq \boldsymbol{0}\). As \(L\) is sufficiently large, the sample covariance matrix $\frac{1}{L} \boldsymbol{X} \boldsymbol{X}^H$ is approximated as the statistical covariance matrix $\boldsymbol{S}$, i.e., \(\frac{1}{L} \boldsymbol{X} \boldsymbol{X}^H \approx \boldsymbol{S}\), which is the optimization variable to be designed.\footnote{This approximation and its accuracy have been verified in the literature of MIMO ISAC systems \cite{liu2021cramer, hua2022mimo}.} Besides, for a maximum transmit power budget \(P\) at the H-AP, we have \(\mathbb{E}\big[\|\boldsymbol{x}(l)\|^2\big] = \mathrm{tr}(\boldsymbol{S}) \le P\).

First, we consider the radar target sensing. The echo signal received by the H-AP receiver is
\begin{equation} \label{Y}
	\boldsymbol{Y}_{\text{S}} = \boldsymbol{H}_{\text{S}} \boldsymbol{X} + \boldsymbol{Z}_{\text{S}},
\end{equation}
where \(\boldsymbol{Z}_{\text{S}} \in \mathbb{C}^{N_{\text{S}} \times L}\) denotes the additive white Gaussian noise (AWGN) at the H-AP receiver that is a CSCG random matrix with independent and identically distributed (i.i.d.) entries each with zero mean and variance \(\sigma_{\text{S}}^2\). The objective of sensing is to estimate the interested target parameters based on \(\boldsymbol{H}_{\text{S}}\) from the received echo signal \(\boldsymbol{Y}_{\text{S}}\) in \eqref{Y}, in which the transmitted signal \(\boldsymbol{X}\) is known by the H-AP. In particular, we consider two different target models, namely the point and extended targets, respectively, as specified in the following.

	For the point target, the target response matrix is
	\begin{equation}
		\boldsymbol{H}_{\text{S}} = \alpha \boldsymbol{a}_r(\theta) \boldsymbol{a}_t^T(\theta) \triangleq \alpha \boldsymbol{A}(\theta), 
	\end{equation}
	where \(\alpha \in \mathbb{C}\) denotes the complex-valued reflection coefficient, \(0 \le \theta \le 2\pi\) denotes the angle of departure/arrival (AoD/AoA), and \(\boldsymbol{a}_t(\theta) \in \mathbb{C}^{M \times 1}\) and \(\boldsymbol{a}_r(\theta) \in \mathbb{C}^{N_{\text{S}} \times 1}\) denote the transmit and receive array steering vectors, respectively. By choosing the center of the ULA antennas as the reference point and assuming half a wavelength spacing between adjacent antennas, it follows that \(\boldsymbol{a}_t(\theta) = \Big[e^{-j \frac{M-1}{2} \pi \sin \theta}, e^{-j \frac{M-3}{2} \pi \sin \theta}, \dots, e^{j \frac{M-1}{2} \pi \sin \theta}\Big]^T\) and \(\boldsymbol{a}_r(\theta) = \Big[e^{-j \frac{N_{\text{S}}-1}{2} \pi \sin \theta}, e^{-j \frac{N_{\text{S}}-3}{2} \pi \sin \theta}, \dots, e^{j \frac{N_{\text{S}}-1}{2} \pi \sin \theta}\Big]^T\).
	In this case, \(\theta\) and \(\alpha\) are the unknown parameters to be estimated. Here, the CRB for estimating \(\theta\) is adopted as the sensing performance metric,\footnote{As acquiring the embedded target information in \(\alpha\) is challenging, in this paper, we focus on the CRB for estimating \(\theta\), similarly as in prior works \cite{liu2021cramer, song2022intelligent}. Besides, if \(\boldsymbol{S}\) is rank-one, then the CRB in \eqref{CRB1} becomes \(\frac{\sigma_{\text{S}}^2}{2|\alpha|^2L \|\boldsymbol{\dot{a}}_r(\theta)\|^2 \boldsymbol{a}_t^T(\theta) \boldsymbol{S} \boldsymbol{a}_t^\dag(\theta)}\) \cite{li2008range}.} which is given by \cite{bekkerman2006target}
	\begin{equation} \label{CRB1}
		\mathrm{CRB}_1(\boldsymbol{S}) = \frac{\frac{\sigma_{\text{S}}^2}{2|\alpha|^2L} \mathrm{tr}(\boldsymbol{A}^H \boldsymbol{A} \boldsymbol{S})}{ \mathrm{tr}(\boldsymbol{\dot{A}}^H \boldsymbol{\dot{A}} \boldsymbol{S}) \mathrm{tr}(\boldsymbol{A}^H \boldsymbol{A} \boldsymbol{S}) -\big|\mathrm{tr}(\boldsymbol{\dot{A}}^H \boldsymbol{A} \boldsymbol{S})\big|^2},
	\end{equation}
	where we define \(\boldsymbol{A} \triangleq \boldsymbol{A}(\theta)\) and \(\boldsymbol{\dot{A}} \triangleq \frac{\partial \boldsymbol{A}(\theta)}{\partial \theta} = \boldsymbol{a}_r(\theta) \boldsymbol{\dot{a}}_t^T(\theta) + \boldsymbol{\dot{a}}_r(\theta) \boldsymbol{a}_t^T(\theta)\). Here, \(\boldsymbol{\dot{a}}_t(\theta)\) and \(\boldsymbol{\dot{a}}_r(\theta)\) denote the derivatives of \(\boldsymbol{a}_t(\theta)\) and \(\boldsymbol{a}_r(\theta)\), respectively, i.e., \(\boldsymbol{\dot{a}}_t(\theta) = \Big[-j a_{t,1} \frac{M-1}{2} \pi \cos \theta, -j a_{t,2} \frac{M-3}{2} \pi \cos \theta, \dots, j a_{t,M} \frac{M-1}{2} \pi \cos \theta\Big]^T\) and \(\boldsymbol{\dot{a}}_r(\theta) = \Big[-j a_{r,1} \frac{N_{\text{S}}-1}{2} \pi \cos \theta, -j a_{r,2} \frac{N_{\text{S}}-3}{2} \pi \cos \theta, \dots,\) \(j a_{r,N_{\text{S}}} \frac{N_{\text{S}}-1}{2} \pi \cos \theta\Big]^T\),
	where \(a_{t,i}\) and \(a_{r,i}\) denote the \(i\)-th entries of \(\boldsymbol{a}_t(\theta)\) and \(\boldsymbol{a}_r(\theta)\), respectively. According to the symmetry of ULA antennas, we have \(\boldsymbol{a}_t^H(\theta) \boldsymbol{\dot{a}}_t(\theta) = 0\) and \(\boldsymbol{a}_r^H(\theta) \boldsymbol{\dot{a}}_r(\theta) = 0\). 
	
	For the extended target, the target response matrix is\cite{liu2021cramer}
	\begin{equation}
		\boldsymbol{H}_{\text{S}} = \sum_{q=1}^Q \alpha_q \boldsymbol{a}_r(\theta_q) \boldsymbol{a}_t^T(\theta_q),
	\end{equation}
	where \(Q\) denotes the number of scatterers at the extended target, \(\alpha_q \in \mathbb{C}\) denotes the complex-valued reflection coefficient associated with the \(q\)-th scatterer, and \(0 \le \theta_{q} \le 2\pi\) denotes the AoD/AoA of the \(q\)-th scatterer. 
	The H-AP aims to estimate the complete target response matrix \(\boldsymbol{H}_{\text{S}}\) as unknown parameters. In this case, the Fisher information matrix (FIM) for estimating \(\boldsymbol{H}_{\text{S}}\) is given by \(\boldsymbol{J} = \frac{1}{\sigma_{\text{S}}^2} \boldsymbol{X}^\dag \boldsymbol{X}^T \otimes \boldsymbol{I}_{N_{\text{S}}} = \frac{L}{\sigma_{\text{S}}^2} \boldsymbol{S}^T \otimes \boldsymbol{I}_{N_{\text{S}}}\), and the corresponding CRB matrix is given by \(\overline{\mathbf{CRB}} = \boldsymbol{J}^{-1}\) \cite{liu2021cramer}. To facilitate performance optimization, we adopt the trace of CRB matrix as a scalar performance metric, i.e.,
	\begin{equation} \label{CRB2}
		\mathrm{CRB}_2(\boldsymbol{S}) = \mathrm{tr}(\overline{\mathbf{CRB}}) =  \frac{\sigma_{\text{S}}^2 N_{\text{S}}}{L} \mathrm{tr}(\boldsymbol{S}^{-1}).
	\end{equation}

Next, we consider the point-to-point MIMO communication from the H-AP to the ID receiver. The received signal by the ID receiver at symbol \(l\) is given by
\begin{equation}
	\boldsymbol{y}_{\text{ID}}(l) = \boldsymbol{H}_{\text{ID}} \boldsymbol{x}(l) + \boldsymbol{z}_{\text{ID}}(l),
\end{equation}
where \(\boldsymbol{z}_{\text{ID}}(l) \in \mathbb{C}^{N_{\text{ID}} \times 1}\) denotes the AWGN at the ID receiver that is a CSCG random vector with zero mean and covariance matrix \(\sigma_{\text{ID}}^2 \boldsymbol{I}_{N_{\text{ID}}}\) and \(\sigma_{\text{ID}}^2\) is the noise power per antenna. With capacity-achieving Gaussian signaling, the achievable data rate (in bps/Hz) is \cite{telatar1999capacity}
\begin{equation} \label{Rate}
	{R}(\boldsymbol{S}) = \log_2 \det \Big(\boldsymbol{I}_{N_{\text{ID}}} + \frac{1}{\sigma_{\text{ID}}^2} \boldsymbol{H}_{\text{ID}} \boldsymbol{S} \boldsymbol{H}_{\text{ID}}^H\Big).
\end{equation}

Then, we consider the WPT from the H-AP to the EH receiver, where the EH receiver adopts rectifiers to convert the received RF signals into DC signals for energy harvesting. In general, the harvested DC power is monotonically non-decreasing with respect to the received RF power. As a result, we exploit the received RF power (energy over a unit time period, in Watt) at the EH receiver as the powering performance metric\footnote{When the RF-to-DC energy conversion process of the rectifier is linear, maximizing the received RF power \({E}(\boldsymbol{S})\) is equivalent to maximizing the harvested DC power. However, maximizing the DC power becomes an interesting and challenging problem when the process is non-linear, which will be left for future work.}  \cite{clerckx2019fundamentals}, i.e.,
\vspace{-0.2cm}\begin{equation} \label{E}
	{E}(\boldsymbol{S}) = \mathrm{tr}(\boldsymbol{H}_{\text{EH}} \boldsymbol{S} \boldsymbol{H}_{\text{EH}}^H).
	\vspace{-0.2cm}
\end{equation}

Our objective is to reveal the fundamental performance tradeoff among the estimation CRB (i.e., \(\mathrm{CRB}_1(\boldsymbol{S})\) in \eqref{CRB1} and \(\mathrm{CRB}_2(\boldsymbol{S})\) in \eqref{CRB2} for the point and extended targets, respectively), the achievable rate \({R}(\boldsymbol{S})\) in \eqref{Rate}, and the received (unit-time) energy \({E}(\boldsymbol{S})\) in \eqref{E}. Towards this end, we aim to characterize the C-R-E region, which is defined as the set of all C-R-E pairs that are simultaneously achievable in the multi-functional wireless MIMO system. For the point target case (\(i=1\)) or the extended case (\(i=2\)), the C-R-E region is defined as
\begin{equation} \label{CRE}
	\begin{aligned}
		\mathcal{C}_i \triangleq \big\{(\widehat{\mathrm{CRB}}_i, \widehat{{R}}, \widehat{{E}}) \big| & \widehat{\mathrm{CRB}}_i \ge \mathrm{CRB}_i(\boldsymbol{S}), \widehat{{R}} \le {R}(\boldsymbol{S}), \\
		& \widehat{{E}} \le {E}(\boldsymbol{S}), \mathrm{tr}(\boldsymbol{S}) \le P, \boldsymbol{S} \succeq \boldsymbol{0}\big\}.
	\end{aligned}
\end{equation}

\section{C-R-E Region Characterization}

This section characterizes the C-R-E region of the multi-functional MIMO system \(\mathcal{C}_i\) in \eqref{CRE}, $i=1,2$, by finding its Pareto boundary, which corresponds to the set of points at which improving one performance metric will inevitably lead to the degradation of another. For instance, Fig. 2 shows the Pareto boundary of an example C-R-E region. It is observed that the complete Pareto boundary corresponds to a surface in a three-dimensional (3D) space, which can be specified by three vertices corresponding to R-max, E-max, and C-min, respectively, as well as three edges corresponding to the optimal C-R, R-E, and C-E tradeoffs, respectively. In the following, we first explain the procedure for acquiring the vertices and edges in Sections \ref{Section:III-A} and \ref{Section:III-B}, respectively, and then formulate the CRB-and-energy-constrained rate maximization problems to determine the remaining interior points on the Pareto boundary surface in Section \ref{Section:III-C}.
\begin{figure}[tb]
	\centering
	{\includegraphics[width=0.30\textwidth]{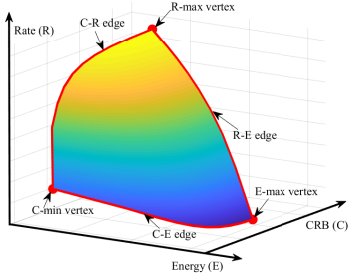}}
	\caption{The Pareto boundary of a C-R-E region example.}
\end{figure}

\subsection{Finding Three Vertices of Pareto Boundary}\label{Section:III-A}

This subsection finds the three vertices on the Pareto boundary of the C-R-E region \(\mathcal{C}_i, \forall i \in \{1,2\}\), which correspond to R-max, E-max, and C-min, respectively. First, we find the R-max vertex by optimizing the transmit covariance matrix $\boldsymbol{S}$ to maximize the communication rate ${R}(\boldsymbol{S})$, i.e., \(\max_{\boldsymbol{S} \succeq \boldsymbol{0}} {R}(\boldsymbol{S}), \ \mathrm{s.t.} \ \mathrm{tr}(\boldsymbol{S}) \le P\), where \(P\) is the maximum transmit power budget. This corresponds to a sole MIMO communication system. It is well-known that the optimal solution to this problem, denoted by \(\boldsymbol{S}_{\text{ID}}\), is obtained by the EMT over the ID channel (i.e., performing SVD to decompose $\boldsymbol{H}_{\text{ID}}$) combined with WF power allocation \cite{telatar1999capacity}. Accordingly, the maximum rate is \({R}_{\mathrm{max}} = {R}(\boldsymbol{S}_{\text{ID}})\) and the corresponding harvested energy and estimation CRB are \({E}_{\text{ID}} = {E}(\boldsymbol{S}_{\text{ID}})\) and \(\mathrm{CRB}_{\text{ID},i} = \mathrm{CRB}_i(\boldsymbol{S}_{\text{ID}}), \forall i \in \{1,2\}\), respectively. As a result, the R-max vertex is obtained as \((\mathrm{CRB}_{\text{ID},i}, {R}_{\mathrm{max}}, {E}_{\text{ID}})\).\footnote{Notice that for the extended target case with $i=2$, if \(\boldsymbol{S}_{\text{ID}}\) is rank-deficient, then we have \(\mathrm{CRB}_{\text{ID},2} = \frac{\sigma_{\text{S}}^2 N_{\text{S}}}{L} \mathrm{tr}(\boldsymbol{S}_{\text{ID}}^{-1}) \to \infty\).}

Next, we obtain the E-max vertex by optimizing $\boldsymbol{S}$ to maximize the harvested energy ${E}(\boldsymbol{S})$, i.e., \(\max_{\boldsymbol{S} \succeq \boldsymbol{0}} {E}(\boldsymbol{S}), \ \mathrm{s.t.} \ \mathrm{tr}(\boldsymbol{S}) \le P\). This corresponds to a sole MIMO WPT system. It has been shown in \cite{zhang2013mimo} that the optimal rank-one solution is obtained based on the strongest EMT, denoted by \(\boldsymbol{S}_{\text{EH}} = \sqrt{P}\boldsymbol{v}_{\text{EH},1} \boldsymbol{v}_{\text{EH},1}^H\), with $\boldsymbol{v}_{\text{EH},1}$ being the dominant right singular vector of $\boldsymbol{H}_{\mathrm{EH}}$. The maximum harvested energy is hence \({E}_{\mathrm{max}} = {E}(\boldsymbol{S}_{\text{EH}})\) and the corresponding communication rate and estimation CRB are \({R}_{\text{EH}} = {R}(\boldsymbol{S}_{\text{EH}})\) and \(\mathrm{CRB}_{\text{EH},i} = \mathrm{CRB}_i(\boldsymbol{S}_{\text{EH}}), \forall i \in \{1,2\}\), respectively. Thus, the E-max vertex is obtained as \((\mathrm{CRB}_{\text{EH},i}, {R}_{\text{EH}}, {E}_{\mathrm{max}})\).\footnote{For the extended target case with $i=2$, \(\mathrm{CRB}_{\text{EH},2} = \frac{\sigma_{\text{S}}^2 N_{\text{S}}}{L} \mathrm{tr}(\boldsymbol{S}_{\text{EH}}^{-1}) \to \infty\) always holds as \(\boldsymbol{S}_{\text{EH}}\) is a rank one matrix.}

Finally, we consider the C-min vertex by optimizing $\boldsymbol{S}$ to minimize the estimation CRB \(\mathrm{CRB}_i(\boldsymbol{S}), \forall i \in \{1,2\}\), 
i.e., \(\min_{\boldsymbol{S} \succeq \boldsymbol{0}} \mathrm{CRB}_i(\boldsymbol{S}), \ \mathrm{s.t.} \ \mathrm{tr}(\boldsymbol{S}) \le P\). This corresponds to a sole MIMO radar sensing system. For the point target case with $i=1$, as derived in \cite{li2008range}, the optimal solution is \(\boldsymbol{S}_{\text{S},1} = \frac{P}{M}\boldsymbol{a}_t^\dag \boldsymbol{a}_t^T\) and the minimum CRB is \(\mathrm{CRB}_{\mathrm{min},1} = \mathrm{CRB}_{1}(\boldsymbol{S}_{\text{S},1}) = \frac{\sigma_{\text{S}}^2}{2 |\alpha|^2 L \|\boldsymbol{\dot{a}}_r\|^2 P M}\).\footnote{Since \(\boldsymbol{S}_{\text{S},1}\) is rank-one, the problem for acquiring the C-min vertex is equivalent to \(\max_{\boldsymbol{S} \succeq \boldsymbol{0}} \boldsymbol{a}_t^T \boldsymbol{S} \boldsymbol{a}_t^\dag, \ \mathrm{s.t.} \ \mathrm{tr}(\boldsymbol{S}) \le P\).} As for the extended target case with $i=2$, the optimal solution is \(\boldsymbol{S}_{\text{S},2} = \frac{P}{M} \boldsymbol{I}_M\) \cite{hua2022mimo}. In this case, the minimum CRB is \(\mathrm{CRB}_{\mathrm{min},2} = \mathrm{CRB}_{2}(\boldsymbol{S}_{\text{S},2}) =  \frac{\sigma_{\text{S}}^2 N_{\text{S}} M^2}{L P}\). With $\boldsymbol{S}_{\text{S},1}$ and $\boldsymbol{S}_{\text{S},2}$ obtained, the corresponding communication rate and harvested energy are \({R}_{\text{S},i} = {R}(\boldsymbol{S}_{\text{S},i})\) and \({E}_{\text{S},i} = {E}(\boldsymbol{S}_{\text{S},i})\), \(\forall i \in \{1,2\}\), respectively. Therefore, we obtain the C-min vertice as \((\mathrm{CRB}_{\mathrm{min},i}, {R}_{\text{S},i}, {E}_{\text{S},i})\).

\subsection{Finding Three Edges of Pareto Boundary}\label{Section:III-B}

This subsection focuses on finding the three edges on the Pareto boundary of the C-R-E region \(\mathcal{C}_i, \forall i \in \{1,2\}\), which capture the optimal C-R, R-E, and C-E tradeoffs, respectively. First, we characterize the C-R edge connecting the C-min vertex \((\mathrm{CRB}_{\mathrm{min},i}, {R}_{\text{S},i}, {E}_{\text{S},i})\) and the R-max vertex \((\mathrm{CRB}_{\text{ID},i}, {R}_{\mathrm{max}}, {E}_{\text{ID}})\) by optimizing $\boldsymbol{S}$ to maximize the communication rate ${R}(\boldsymbol{S})$ subject to a maximum CRB constraint, i.e., 
\(\max_{\boldsymbol{S} \succeq \boldsymbol{0}} {R}(\boldsymbol{S}), \ \mathrm{s.t.} \ \mathrm{CRB}_i(\boldsymbol{S}) \le \mathrm{CRB}_{\text{C-R},i}, \ \mathrm{tr}(\boldsymbol{S}) \le P\), where the CRB threshold \(\mathrm{CRB}_{\text{C-R},i}\) ranges from \(\mathrm{CRB}_{\mathrm{min},i}\) to \(\mathrm{CRB}_{\text{ID},i}\) to obtain different C-R tradeoff points on the edge. This corresponds to the CRB-constrained rate maximization problem for a MIMO ISAC system without WPT, whose optimal solution has been derived in \cite{hua2022mimo} in a semi-closed-form, denoted by \(\boldsymbol{S}_{\text{C-R},i}\). Let \({R}_{\text{C-R},i}(\mathrm{CRB}_{\text{C-R},i}) = {R}(\boldsymbol{S}_{\text{C-R},i})\) denote the achieved rate and \({E}_{\text{C-R},i}(\mathrm{CRB}_{\text{C-R},i}) = {E}(\boldsymbol{S}_{\text{C-R},i})\) denote the corresponding harvested energy. Then, the C-R tradeoff edge for case \(i \in \{1,2\}\) is \(\Big\{\big(\mathrm{CRB}_{\text{C-R},i}, {R}_{\text{C-R},i}(\mathrm{CRB}_{\text{C-R},i}), {E}_{\text{C-R},i}(\mathrm{CRB}_{\text{C-R},i})\big) \big| \mathrm{CRB}_{\mathrm{min},i} \le \mathrm{CRB}_{\text{C-R},i} \le \mathrm{CRB}_{\text{ID},i}\Big\}\).

Next, we obtain the R-E edge connecting the R-max vertex \((\mathrm{CRB}_{\text{ID},i}, {R}_{\mathrm{max}}, {E}_{\text{ID}})\) and the E-max vertex \((\mathrm{CRB}_{\text{EH},i}, {R}_{\text{EH}}, {E}_{\mathrm{max}})\) by optimizing $\boldsymbol{S}$ to maximize the rate ${R}(\boldsymbol{S})$ subject to a minimum energy constraint, i.e., \(\max_{\boldsymbol{S} \succeq \boldsymbol{0}} {R}(\boldsymbol{S}), \ \mathrm{s.t.} \ {E}(\boldsymbol{S}) \ge {E}_{\text{R-E}}, \ \mathrm{tr}(\boldsymbol{S}) \le P\), where the energy threshold \({E}_{\text{R-E}}\) ranges from \({E}_{\text{ID}}\) to \({E}_{\mathrm{max}}\) for searching different R-E tradeoff points on the edge. This corresponds to the energy-constrained rate maximization problem for a MIMO SWIPT system without radar sensing, whose optimal transmit covariance has been obtained in \cite{zhang2013mimo} as \(\boldsymbol{S}_{\text{R-E}}\) in a semi-closed-form. Accordingly, let \({R}_{\text{R-E}}({E}_{\text{R-E}}) = {R}(\boldsymbol{S}_{\text{R-E}})\) denote the achieved rate and \(\mathrm{CRB}_{\text{R-E},i}({E}_{\text{R-E}}) = \mathrm{CRB}_i(\boldsymbol{S}_{\text{R-E}})\) denote the corresponding CRBs. Thus, the R-E tradeoff edge for case \(i \in \{1,2\}\) is given by \(\Big\{\big(\mathrm{CRB}_{\text{R-E},i}({E}_{\text{R-E}}), {R}_{\text{R-E}}({E}_{\text{R-E}}), {E}_{\text{R-E}}\big) | {E}_{\text{ID}} \le {E}_{\text{R-E}} \le {E}_{\mathrm{max}}\Big\}\).

Finally, we consider the C-E edge to connect the C-min vertex \((\mathrm{CRB}_{\mathrm{min},i}, {R}_{\text{S},i}, {E}_{\text{S},i})\) and the E-max vertex \((\mathrm{CRB}_{\text{EH},i}, {R}_{\text{EH}}, {E}_{\mathrm{max}})\) by optimizing $\boldsymbol{S}$ to maximize the energy ${E}(\boldsymbol{S})$ subject to a maximum CRB constraint, i.e., \(\max_{\boldsymbol{S} \succeq \boldsymbol{0}} {E}(\boldsymbol{S}), \ \mathrm{s.t.} \ \mathrm{CRB}_i(\boldsymbol{S}) \le \mathrm{CRB}_{\text{C-E},i}, \ \mathrm{tr}(\boldsymbol{S}) \le P\), where the CRB threshold \(\mathrm{CRB}_{\text{C-E},i}\) ranges from \(\mathrm{CRB}_{\mathrm{min},i}\) to \(\mathrm{CRB}_{\text{EH},i}\) to obtain different C-E tradeoff points on the edge, which corresponds to the case when only sensing and WPT are considered in our integrated system. For the point target case with \(i=1\), the optimization problem is reformulated into a convex form, as the CRB constraint \(\mathrm{CRB}_1(\boldsymbol{S}) \le \mathrm{CRB}_{\text{C-E},1}\) is equivalent to the convex semi-definite constraint \(\begin{bmatrix}
	\mathrm{tr}(\boldsymbol{\dot{A}}^H \boldsymbol{\dot{A}} \boldsymbol{S}) - \frac{\sigma_{\text{S}}^2}{2 |\alpha|^2 L \mathrm{CRB}_{\text{C-E},1}} & \mathrm{tr}(\boldsymbol{\dot{A}}^H \boldsymbol{A} \boldsymbol{S})^\dag \\
	\mathrm{tr}(\boldsymbol{\dot{A}}^H \boldsymbol{A} \boldsymbol{S}) & \mathrm{tr}(\boldsymbol{A}^H \boldsymbol{A} \boldsymbol{S})
\end{bmatrix} \succeq \boldsymbol{0}\) based on the Schur complement \cite{liu2021cramer}. Therefore, the optimal solution \(\boldsymbol{S}_{\text{C-E},1}\) in this case can be found by standard convex optimization techniques \cite{grant2014cvx}.\footnote{Note that we can obtain an analytical optimal solution to this problem by using Lagrangian duality method. However, we omit the derivation here as it is the subcase of problem (P1) in Section IV, where we will show that \(\boldsymbol{S}_{\text{C-E},1}\) is rank-one and this problem is equivalent to \(\max_{\boldsymbol{S} \succeq \boldsymbol{0}} E(\boldsymbol{S}), \ \mathrm{s.t.} \ \boldsymbol{a}_t^T \boldsymbol{S} \boldsymbol{a}_t^\dag \ge \frac{\sigma_{\text{S}}^2}{2 |\alpha|^2 L \|\boldsymbol{\dot{a}}_r\|^2 \mathrm{CRB}_{\text{C-E},1}}, \ \mathrm{tr}(\boldsymbol{S}) \le P\).} For the extended target case with \(i=2\), the semi-closed-form optimal solution \(\boldsymbol{S}_{\text{C-E},2}\) is obtained similarly as in \cite{ren2022fundamental}. Let \({E}_{\text{C-E},i}(\mathrm{CRB}_{\text{C-E},i}) = {E}(\boldsymbol{S}_{\text{C-E},i})\) denote the achieved harvested energy and \({R}_{\text{C-E},i}(\mathrm{CRB}_{\text{C-E},i}) = {R}(\boldsymbol{S}_{\text{C-E},i})\) denote the corresponding communication rate. Accordingly, the C-E tradeoff edge for case \(i \in \{1,2\}\) is \(\Big\{\big(\mathrm{CRB}_{\text{C-E},i}, {R}_{\text{C-E},i}(\mathrm{CRB}_{\text{C-E},i}), {E}_{\text{C-E},i}(\mathrm{CRB}_{\text{C-E},i})\big) \big| \mathrm{CRB}_{\mathrm{min},i} \le \mathrm{CRB}_{\text{C-E},i} \le \mathrm{CRB}_{\text{EH},i}\Big\}\).

\subsection{Characterizing Complete Pareto Boundary Surface}\label{Section:III-C}

With the vertices and edges obtained, it remains to find the interior points on the Pareto boundary surface to characterize the complete C-R-E region \(\mathcal{C}_i\) for case $i \in \{1,2\}$. Towards this end, we optimize the transmit covariance $\boldsymbol{S}$ to maximize the communication rate ${R}(\boldsymbol{S})$ in \eqref{Rate}, subject to a maximum estimation CRB constraint $\Gamma_{\text{S}}$ and a minimum harvested energy constraint $\Gamma_{\text{EH}}$. Mathematically, the CRB-and-energy-constrained rate maximization problem for case $i \in \{1,2\}$ is formulated as 
\begin{subequations}
	\begin{align} 
		(\text{P}i): \max_{\boldsymbol{S} \succeq \boldsymbol{0}} &\ \log_2 \det \Big(\boldsymbol{I}_{N_{\text{ID}}} + \frac{1}{\sigma_{\text{ID}}^2} \boldsymbol{H}_{\text{ID}} \boldsymbol{S} \boldsymbol{H}_{\text{ID}}^H\Big) \label{R} \\
		\mathrm{s.t.} &\ \mathrm{tr}(\boldsymbol{H}_{\text{EH}} \boldsymbol{S} \boldsymbol{H}_{\text{EH}}^H) \ge \Gamma_{\text{EH}}, \label{Ra} \\
		&\ \mathrm{CRB}_i(\boldsymbol{S}) \le \Gamma_{\text{S}}, \label{Rb} \\ 
		&\ \mathrm{tr}(\boldsymbol{S}) \le P. \label{Rc}
	\end{align}
\end{subequations}
By exhausting the possible values of \(\Gamma_{\text{EH}}\) and \(\Gamma_{\text{S}}\) enclosed by the projection of the three edges on the C-E plane, we can obtain all the Pareto boundary surface points \cite{zhang2013mimo, hua2022mimo}. In particular, let \(R_{i}^\star(\Gamma_{\text{EH}}, \Gamma_{\text{S}})\) denote the optimal value of problem (P\(i\)) with given \(\Gamma_{\text{EH}}\) and \(\Gamma_{\text{S}}\). Then, we have one Pareto boundary point of the C-R-E region \(\mathcal{C}_i\) as \(\big(\Gamma_{\text{S}}, R_{i}^\star(\Gamma_{\text{EH}}, \Gamma_{\text{S}}), \Gamma_{\text{EH}}\big)\).  

In the sequel, we focus on solving problems (P1) and (P2) for the point and extended target cases, respectively.

\section{Optimal Solution to Problem (P1) with Point Target}

In this section, we present the optimal solution to problem (P1) in the point target case. Notice that (P1) is not convex due to the estimation CRB constraint in \eqref{Rb} for $i=1$. Fortunately, based on the Schur complement, constraint \eqref{Rb} for $i=1$ is equivalent to
\begin{equation}\label{Rb-reformulation}
	\begin{bmatrix}
		\mathrm{tr}(\boldsymbol{\dot{A}}^H \boldsymbol{\dot{A}} \boldsymbol{S}) - \frac{1}{\Gamma_{\text{S},1}} & \mathrm{tr}(\boldsymbol{\dot{A}}^H \boldsymbol{A} \boldsymbol{S})^\dag \\
		\mathrm{tr}(\boldsymbol{\dot{A}}^H \boldsymbol{A} \boldsymbol{S}) & \mathrm{tr}(\boldsymbol{A}^H \boldsymbol{A} \boldsymbol{S})
	\end{bmatrix} \succeq \boldsymbol{0},
\end{equation}
where \(\Gamma_{\text{S},1} \triangleq \frac{2|\alpha|^2L}{\sigma_{\text{S}}^2} \Gamma_{\text{S}}\). Therefore, by replacing constraint \eqref{Rb} with \eqref{Rb-reformulation}, we reformulate (P1) into a convex form and optimally solve it by adopting the Lagrangian duality method that paves the way for analyzing the structure of the optimal transmit covariance matrix.

Let \(\lambda \ge 0\), \(\nu \ge 0\), and \(\boldsymbol{Z} \triangleq \begin{bmatrix}
	z_1 & z_2 \\ z_2^\dag & z_3
\end{bmatrix} \succeq \boldsymbol{0}\) denote the dual variables associated with the constraints in \eqref{Ra}, \eqref{Rc}, and \eqref{Rb-reformulation}, respectively. The Lagrangian of problem (P1) is given by 
\begin{equation}
	\begin{aligned}
		\mathcal{L}(\boldsymbol{S}, \lambda, \nu, \boldsymbol{Z}) =&
		 \log_2 \det \Big(\boldsymbol{I}_{N_{\text{ID}}} + \frac{1}{\sigma_{\text{ID}}^2} \boldsymbol{H}_{\text{ID}} \boldsymbol{S} \boldsymbol{H}_{\text{ID}}^H\Big) - \mathrm{tr}(\boldsymbol{D} \boldsymbol{S}) \\
		 & - \lambda \Gamma_{\text{EH}} + \nu P - \frac{z_1}{\Gamma_{\text{S},1}},
	\end{aligned}
\end{equation}
where \(\boldsymbol{D} \triangleq \nu \boldsymbol{I}_M - \lambda \boldsymbol{H}_{\text{EH}}^H \boldsymbol{H}_{\text{EH}} -  z_1 \boldsymbol{\dot{A}}^H \boldsymbol{\dot{A}} - z_2 \boldsymbol{\dot{A}}^H \boldsymbol{A} - z_2^\dag \boldsymbol{A}^H \boldsymbol{\dot{A}} - z_3 \boldsymbol{A}^H \boldsymbol{A}\). Note that since \(\boldsymbol{\dot{A}} = \boldsymbol{a}_r \boldsymbol{\dot{a}}_t^T + \boldsymbol{\dot{a}}_r \boldsymbol{a}_t^T\) and \(\boldsymbol{a}_t^H \boldsymbol{\dot{a}}_t = \boldsymbol{a}_r^H \boldsymbol{\dot{a}}_r = 0\), it follows that \(\boldsymbol{D} = \nu \boldsymbol{I}_M - \lambda \boldsymbol{H}_{\text{EH}}^H \boldsymbol{H}_{\text{EH}} - (z_1 \boldsymbol{\dot{a}}_t^\dag \boldsymbol{\dot{a}}_t^T + z_2 \boldsymbol{\dot{a}}_t^\dag \boldsymbol{a}_t^T + z_2^\dag \boldsymbol{a}_t^\dag \boldsymbol{\dot{a}}_t^T + z_3 \boldsymbol{a}_t^\dag \boldsymbol{a}_t^T) \|\boldsymbol{a}_r\|^2 - z_1 \boldsymbol{a}_t^\dag \boldsymbol{a}_t^T \|\boldsymbol{\dot{a}}_r\|^2\).

Accordingly, the dual function of (P1) is defined as 
\begin{equation} \label{dualfunc_p}
	g(\lambda, \nu, \boldsymbol{Z}) = \max_{\boldsymbol{S} \succeq \boldsymbol{0}} \ \mathcal{L}(\boldsymbol{S},\lambda, \nu, \boldsymbol{Z}).
\end{equation}
We have the following lemma, which can be verified similarly as in \cite[Lemma 1]{hua2022mimo}.  

\begin{lemma} \label{lem1}
	If the dual function \(g(\lambda, \nu, \boldsymbol{Z})\) is bounded from above, then it holds that 
	\begin{equation}\label{D1-constraint}
		\boldsymbol{D} \succeq \boldsymbol{0}~{\text{and}}~\mathcal{R}(\boldsymbol{H}_{\text{ID}}^H) \subseteq \mathcal{R}(\boldsymbol{D}).
	\end{equation}
\end{lemma}

Based on Lemma \ref{lem1}, the dual problem of (P1) is defined as 
\begin{equation}
	(\text{D}1): \min_{\lambda \ge 0, \nu \ge 0, \boldsymbol{Z} \succeq \boldsymbol{0}} \ g(\lambda, \nu, \boldsymbol{Z}), \ \mathrm{s.t.} \ \eqref{D1-constraint}.
\end{equation}

Since problem (P1) is reformulated into a convex form and satisfies the Slater's condition, strong duality holds between (P1) and its dual problem (D1) \cite{boyd2004vandenberghe}. Therefore, we can solve (P1) by equivalently solving (D1). In the following, we first obtain the dual function \(g(\lambda, \nu, \boldsymbol{Z})\) with given dual variables, and then search over \(\lambda \ge 0\), \(\nu \ge 0\), and \(\boldsymbol{Z} \succeq \boldsymbol{0}\) to minimize \(g(\lambda, \nu, \boldsymbol{Z})\).

\subsection{Finding Dual Function \(g(\lambda, \nu, \boldsymbol{Z})\) of Problem (P1)}

First, we find the dual function \(g(\lambda, \nu, \boldsymbol{Z})\) in \eqref{dualfunc_p} for given \(\lambda \ge 0\), \(\nu \ge 0\), and \(\boldsymbol{Z} \succeq \boldsymbol{0}\)  satisfying \eqref{D1-constraint}. By dropping the constant terms, the problem in \eqref{dualfunc_p} is equivalent to
\begin{equation} \label{L_P_1}
	\max_{\boldsymbol{S} \succeq \boldsymbol{0}} \ \log_2 \det \Big(\boldsymbol{I}_{N_{\text{ID}}} + \frac{1}{\sigma_{\text{ID}}^2} \boldsymbol{H}_{\text{ID}} \boldsymbol{S} \boldsymbol{H}_{\text{ID}}^H\Big) - \mathrm{tr}(\boldsymbol{D} \boldsymbol{S}).
\end{equation}

Suppose that \(\mathrm{rank}(\boldsymbol{D}) = r_{\text{p}}\), the eigenvalue decomposition (EVD) of \(\boldsymbol{D}\) can be expressed as
\begin{equation} \label{D}
	\boldsymbol{D} = 
	\begin{bmatrix}
		\boldsymbol{Q}^{\overline{\mathrm{null}}} \ \boldsymbol{Q}^{\mathrm{null}}
	\end{bmatrix}
	\begin{bmatrix}
		\boldsymbol{\Sigma}^{\overline{\mathrm{null}}} & \boldsymbol{0} \\ \boldsymbol{0} & \boldsymbol{0}
	\end{bmatrix}
	\begin{bmatrix}
		{\boldsymbol{Q}^{\overline{\mathrm{null}}}}^H \\ {\boldsymbol{Q}^{\mathrm{null}}}^H
	\end{bmatrix},
\end{equation}
where \(\boldsymbol{\Sigma}^{\overline{\mathrm{null}}} = \mathrm{diag}(\sigma_{\text{p},1}, \dots, \sigma_{\text{p},r_{\text{p}}})\), and \({\boldsymbol{Q}^{\overline{\mathrm{null}}}} \in \mathbb{C}^{M \times r_{\text{p}}}\) and \({\boldsymbol{Q}^{\mathrm{null}}} \in \mathbb{C}^{M \times (M-r_{\text{p}})}\) denote the eigenvectors corresponding to the \(r_{\text{p}}\) non-zero eigenvalues \(\sigma_{\text{p},1}, \dots, \sigma_{\text{p},r_{\text{p}}}\), and the remaining \(M-r_{\text{p}}\) zero eigenvalues, respectively. Without loss of generality, we express \(\boldsymbol{S}\) as
\begin{equation} \label{S_p}
		\boldsymbol{S} = \begin{bmatrix}
			\boldsymbol{Q}^{\overline{\mathrm{null}}} \ \boldsymbol{Q}^{\mathrm{null}}
		\end{bmatrix}
		\begin{bmatrix}
			\boldsymbol{S}_{11} & \boldsymbol{C} \\ \boldsymbol{C}^H & \boldsymbol{S}_{00}
		\end{bmatrix}
		\begin{bmatrix}
			{\boldsymbol{Q}^{\overline{\mathrm{null}}}}^H \\ {\boldsymbol{Q}^{\mathrm{null}}}^H
		\end{bmatrix}, 
\end{equation}
where \(\boldsymbol{S}_{11} \in \mathbb{S}^{r_{\text{p}}}\), \(\boldsymbol{S}_{00} \in \mathbb{S}^{M-r_{\text{p}}}\), and \(\boldsymbol{C} \in \mathbb{C}^{r_{\text{p}} \times (M-r_{\text{p}})}\) are variables to be optimized. We define \(\mathrm{rank}\big(\boldsymbol{H}_{\text{ID}} \boldsymbol{Q}^{\overline{\mathrm{null}}} (\boldsymbol{\Sigma}^{\overline{\mathrm{null}}})^{-\frac{1}{2}}\big) = \widetilde{r}_{\text{p}}\) and the truncated SVD \(\boldsymbol{H}_{\text{ID}} \boldsymbol{Q}^{\overline{\mathrm{null}}} (\boldsymbol{\Sigma}^{\overline{\mathrm{null}}})^{-\frac{1}{2}} = \boldsymbol{U}_{\text{p}} \boldsymbol{\Lambda}_{\text{p}} \boldsymbol{V}_{\text{p}}^H\), where \(\boldsymbol{U}_{\text{p}} \in \mathbb{C}^{N_{\text{ID}} \times \widetilde{r}_{\text{p}}}\), \(\boldsymbol{V}_{\text{p}} \in \mathbb{C}^{r_{\text{p}} \times \widetilde{r}_{\text{p}}}\), and \(\boldsymbol{\Lambda}_{\text{p}} = \mathrm{diag}(\lambda_{\text{p},1}, \dots, \lambda_{\text{p},\widetilde{r}_{\text{p}}})\). We have the following result.

\begin{proposition} \label{prop1}
	The optimal solution to problem \eqref{L_P_1} is
	\begin{equation} \label{22}
		\boldsymbol{S}^* = \begin{bmatrix}
			\boldsymbol{Q}^{\overline{\mathrm{null}}} \ \boldsymbol{Q}^{\mathrm{null}}
		\end{bmatrix} \begin{bmatrix}
			\boldsymbol{S}_{11}^* & \boldsymbol{C}^* \\ {\boldsymbol{C}^*}^H & \boldsymbol{S}_{00}^*
		\end{bmatrix} \begin{bmatrix}
			{\boldsymbol{Q}^{\overline{\mathrm{null}}}}^H \\ {\boldsymbol{Q}^{\mathrm{null}}}^H
		\end{bmatrix},
	\end{equation}
	where
	\begin{equation} \label{S_11}
		\boldsymbol{S}_{11}^* = (\boldsymbol{\Sigma}^{\overline{\mathrm{null}}})^{-\frac{1}{2}} \boldsymbol{V}_{\text{p}} \mathrm{diag}(\widetilde{p}_1, \dots, \widetilde{p}_{\widetilde{r}_{\text{p}}}) \boldsymbol{V}_{\text{p}}^H (\boldsymbol{\Sigma}^{\overline{\mathrm{null}}})^{-\frac{1}{2}},
	\end{equation}
	with
	\begin{equation} \label{power_allo}
		\widetilde{p}_k = \Big(\frac{1}{\ln 2}-\frac{\sigma_{\text{ID}}^2}{\lambda_{\text{p},k}^2}\Big)^+, \forall k \in \{1,\dots,\widetilde{r}_{\text{p}}\},
	\end{equation}
	and \(\boldsymbol{C}^*\) and \(\boldsymbol{S}_{00}^* \succeq \boldsymbol{0}\) can be chosen arbitrarily such that \(\boldsymbol{S}^* \succeq \boldsymbol{0}\).\footnote{Note that \(\boldsymbol{C}^*\) and \(\boldsymbol{S}_{00}^*\) are generally non-unique here. As a result, an additional step is needed to determine them for solving the primal problem (P1) later. Here, we can simply choose \(\boldsymbol{C}^* = \boldsymbol{0}\) and \(\boldsymbol{S}_{00}^* = \boldsymbol{0}\) for obtaining the dual function \(g(\lambda,\nu,\boldsymbol{Z})\) only.}
\end{proposition}
	
\textit{Proof:} See Appendix A. 
\hfill \(\square\)

\subsection{Solving Dual Problem (D1)}

Next, we solve the dual problem (D1), which is convex but not necessarily differentiable. Therefore, we can solve (D1) by applying subgradient-based methods such as the ellipsoid method \cite{boyd2004vandenberghe}. First, for the objective function \(g(\lambda, \nu, \boldsymbol{Z})\), the subgradient at \((\lambda, \nu, z_1, z_2, z_3)\) is \(\big[\mathrm{tr}(\boldsymbol{H}_{\text{EH}}^H \boldsymbol{H}_{\text{EH}} \boldsymbol{S}^*) - \Gamma_{\text{EH}},
P - \mathrm{tr}(\boldsymbol{S}^*),
\mathrm{tr}(\boldsymbol{\dot{A}}^H \boldsymbol{\dot{A}} \boldsymbol{S}^*) - \frac{1}{\Gamma_{\text{S},1}},
\mathrm{tr}(\boldsymbol{\dot{A}}^H \boldsymbol{A} \boldsymbol{S}^* + \boldsymbol{A}^H \boldsymbol{\dot{A}} \boldsymbol{S}^*) + j \mathrm{tr}(\boldsymbol{\dot{A}}^H \boldsymbol{A} \boldsymbol{S}^* - \boldsymbol{A}^H \boldsymbol{\dot{A}} \boldsymbol{S}^*),
\mathrm{tr}(\boldsymbol{A}^H \boldsymbol{A} \boldsymbol{S}^*)\big]^T\).
Then, let \(\boldsymbol{q}_1\) denote the eigenvector corresponding to the minimum eigenvalue of \(\boldsymbol{D}\). Since constraint \(\boldsymbol{D} \succeq \boldsymbol{0}\) is equivalent to \(\boldsymbol{q}_1^H \boldsymbol{D} \boldsymbol{q}_1 \ge 0\), the subgradient of constraint \(\boldsymbol{D} \succeq \boldsymbol{0}\) at \((\lambda, \nu, z_1, z_2, z_3)\) is \(\big[\boldsymbol{q}_1^H \boldsymbol{H}_{\text{EH}}^H \boldsymbol{H}_{\text{EH}} \boldsymbol{q}_1,
-1,
\boldsymbol{q}_1^H \boldsymbol{\dot{A}}^H \boldsymbol{\dot{A}} \boldsymbol{q}_1,
\boldsymbol{q}_1^H (\boldsymbol{\dot{A}}^H \boldsymbol{A} + \boldsymbol{A}^H \boldsymbol{\dot{A}}) \boldsymbol{q}_1 + j \boldsymbol{q}_1^H (\boldsymbol{\dot{A}}^H \boldsymbol{A} - \boldsymbol{A}^H \boldsymbol{\dot{A}}) \boldsymbol{q}_1,
\boldsymbol{q}_1^H \boldsymbol{A}^H \boldsymbol{A} \boldsymbol{q}_1\big]^T\).
Furthermore, let \(\boldsymbol{q}_2 = [q_{2,1}, q_{2,2}]^T\) denote the eigenvector corresponding to the minimum eigenvalue of \(\boldsymbol{Z}\). The subgradient of constraint \(\boldsymbol{Z} \succeq \boldsymbol{0}\) at \((\lambda, \nu, z_1, z_2, z_3)\) is \(\big[0,
0,
-|q_{2,1}|^2,
-(q_{2,1}^\dag q_{2,2} + q_{2,2}^\dag q_{2,1}) - j (q_{2,1}^\dag q_{2,2} - q_{2,2}^\dag q_{2,1}),
-|q_{2,2}|^2\big]^T\). With these derived subgradients, the ellipsoid method can be implemented efficiently, based on which we can obtain the optimal dual solution to (D1) as \(\lambda^*\), \(\nu^*\), and \(\boldsymbol{Z}^*\).

\subsection{Optimal Solution to Primal Problem (P1)}

Now, we present the optimal solution to the primal problem (P1). With the optimal dual variables \(\lambda^*\), \(\nu^*\), and \(\boldsymbol{Z}^*\) at hand, the corresponding unique optimal solution \(\boldsymbol{S}_{11}^*\) to problem \eqref{L_P_1} is directly used for constructing the optimal primal solution to (P1), denoted by \(\boldsymbol{S}_{11}^{\mathrm{opt}}\). However, as indicated in Section IV-A, the optimal solutions of \(\boldsymbol{C}^*\) and \(\boldsymbol{S}_{00}^*\) to \eqref{L_P_1} are not unique. As a result, with given \(\boldsymbol{S}_{11}^{\mathrm{opt}}\), we need to obtain the optimal solutions of \(\boldsymbol{C}\) and \(\boldsymbol{S}_{00}\), denoted by \(\boldsymbol{C}^{\mathrm{opt}}\) and \(\boldsymbol{S}_{00}^{\mathrm{opt}}\), respectively, by exploiting the following proposition.

\begin{proposition} \label{prop2}
	The optimal solution to problem (P1) is
	\begin{equation} \label{P1_opt}
		\boldsymbol{S}^{\mathrm{opt},1} = \begin{bmatrix}
			\boldsymbol{Q}^{\overline{\mathrm{null}}} \ \boldsymbol{Q}^{\mathrm{null}}
		\end{bmatrix} \begin{bmatrix}
			\boldsymbol{S}_{11}^{\mathrm{opt}} & \boldsymbol{C}^{\mathrm{opt}} \\ {\boldsymbol{C}^{\mathrm{opt}}}^H & \boldsymbol{S}_{00}^{\mathrm{opt}}
		\end{bmatrix} \begin{bmatrix}
			{\boldsymbol{Q}^{\overline{\mathrm{null}}}}^H \\ {\boldsymbol{Q}^{\mathrm{null}}}^H
		\end{bmatrix},
	\end{equation}
	where \(\boldsymbol{S}_{11}^{\mathrm{opt}}\) is given by \eqref{S_11} based on \(\lambda^*\), \(\nu^*\), and \(\boldsymbol{Z}^*\), while \(\boldsymbol{C}^{\mathrm{opt}}\) and \(\boldsymbol{S}_{00}^{\mathrm{opt}}\) are obtained by solving the following feasibility problem.
	\begin{equation}
		\begin{aligned}
			\mathrm{find} \ \boldsymbol{C}, \boldsymbol{S}_{00}, \
			& \mathrm{s.t.} \ \eqref{Ra}, \ \eqref{Rc}, \ \eqref{Rb-reformulation}, \\ &\ \ \ \boldsymbol{S} = \begin{bmatrix}
				\boldsymbol{Q}^{\overline{\mathrm{null}}} \ \boldsymbol{Q}^{\mathrm{null}}
			\end{bmatrix} \begin{bmatrix}
				\boldsymbol{S}_{11}^{\mathrm{opt}} & \boldsymbol{C} \\ \boldsymbol{C}^H & \boldsymbol{S}_{00}
			\end{bmatrix} \begin{bmatrix}
				{\boldsymbol{Q}^{\overline{\mathrm{null}}}}^H \\ {\boldsymbol{Q}^{\mathrm{null}}}^H
			\end{bmatrix}.
		\end{aligned}
	\end{equation}
\end{proposition}

\begin{remark} \label{rem1}
	Based on Proposition \ref{prop2}, we have the following interesting observations on the optimal transmit covariance matrix \(\boldsymbol{S}^{\mathrm{opt},1}\) for the point target case. First, it is observed that \(\boldsymbol{S}^{\mathrm{opt},1}\) can be divided into two parts, i.e., \(\boldsymbol{S}_{11}^{\mathrm{opt}}\) for the triple roles of communication, sensing, and powering; and \(\boldsymbol{C}^{\mathrm{opt}}\) and \(\boldsymbol{S}_{00}^{\mathrm{opt}}\) for sensing and WPT only. Next, due to the competition of the three functionalities,\footnote{From \eqref{22} and \eqref{S_11}, \(\boldsymbol{S}^{\mathrm{opt,1}}\) can be decomposed as \(\boldsymbol{S}^{\mathrm{opt,1}} = \boldsymbol{F} \boldsymbol{\Xi} \boldsymbol{F}^H\). Here, \(\boldsymbol{\Xi}\) is a positive semidefinite matrix, while \(\boldsymbol{F}\) is a semi-unitary matrix whose columns are a set of orthogonal uniform basis of \(\mathrm{span}\big\{\boldsymbol{a}_t^\dag, \boldsymbol{\dot{a}}_t^\dag, \{\boldsymbol{v}_{\text{ID},i}\}, \{\boldsymbol{v}_{\text{EH},i}\}\big\}\). Here, \(\boldsymbol{v}_{\text{ID},i}\)'s and \(\boldsymbol{v}_{\text{EH},i}\)'s are the right singular vectors corresponding to the non-zero singular values of \(\boldsymbol{H}_{\text{ID}}\) and \(\boldsymbol{H}_{\text{EH}}\), respectively.} \(\boldsymbol{S}_{11}^{\mathrm{opt}}\) follows the EMT structure based on the ID channel \(\boldsymbol{H}_{\text{ID}}\) and the composite channel matrix \(\boldsymbol{D}\) consisting of sensing and EH channels (cf. \eqref{S_11}), together with a WF-like power allocation (cf. \eqref{power_allo}). Here, the matrix \(\boldsymbol{D}\) properly balances the tradeoff between sensing and WPT, and the EMT structure of \(\boldsymbol{S}_{11}^{\mathrm{opt}}\) highlights the tradeoff between the communication versus the combined sensing and WPT. Therefore, the obtained solution provides an efficient tradeoff between ISAC and SWIPT.
	Notice that based on extensive simulations, we find that with randomly generated channels, it follows that \(\boldsymbol{D} \succ \boldsymbol{0}\) and \(\boldsymbol{S}^{\mathrm{opt},1} = \boldsymbol{Q}^{\overline{\mathrm{null}}} \boldsymbol{S}_{11}^{\mathrm{opt}} {\boldsymbol{Q}^{\overline{\mathrm{null}}}}^H\), i.e., only \(\boldsymbol{S}_{11}^{\mathrm{opt}}\) is necessary; by contrast, in some special cases (e.g., \(\boldsymbol{H}_{\text{ID}}\) and \(\boldsymbol{H}_{\text{EH}}\) being orthogonal to \(\boldsymbol{A}\)), it could happen that \(\boldsymbol{D}\) is rank-deficient, and thus \(\boldsymbol{C}^{\mathrm{opt}}\) and \(\boldsymbol{S}_{00}^{\mathrm{opt}}\) are also required.
\end{remark}

\subsection{Special Cases with Co-located C-R and C-E}

\subsubsection{Co-located C-R setup}

In this setup, we perform SWIPT for separate ID and EH receivers, while locating the ID receiver concurrently. Accordingly, the LoS communication channel is \(\boldsymbol{H}_{\text{ID}} = \alpha_{\text{ID}} \boldsymbol{a}_{\text{ID}}(\theta) \boldsymbol{a}_t^T(\theta)\), where \(\alpha_{\text{ID}}\) is the path loss and \(\boldsymbol{a}_{\text{ID}}(\theta) = \Big[e^{-j \frac{N_{\text{ID}}-1}{2} \pi \sin \theta}, e^{-j \frac{N_{\text{ID}}-3}{2} \pi \sin \theta}, \dots, e^{j \frac{N_{\text{ID}}-1}{2} \pi \sin \theta}\Big]^T\) is the steering vector at the ID receiver.
Since \(\boldsymbol{a}_{\text{ID}}^H \boldsymbol{a}_{\text{ID}} = N_{\text{ID}}\) and the communication rate is \(R(\boldsymbol{S}) = \log_2 \Big(1 + \frac{\alpha_{\text{ID}}^2 N_{\text{ID}}}{\sigma_{\text{ID}}^2} \boldsymbol{a}_t^T \boldsymbol{S} \boldsymbol{a}_t^\dag\Big)\), we can treat \(\alpha_{\text{ID}} \sqrt{N_{\text{ID}}} \boldsymbol{a}_t^T\) as the equivalent ID channel. According to Proposition 1, the optimal solution to (P1) in this case is rank-one, and thus (P1) is equivalent to
\begin{equation} \label{31}
	\max_{\boldsymbol{S} \succeq \boldsymbol{0}} \ \boldsymbol{a}_t^T \boldsymbol{S} \boldsymbol{a}_t^\dag, \ \mathrm{s.t.} \ \mathrm{tr}(\boldsymbol{H}_{\text{EH}} \boldsymbol{S} \boldsymbol{H}_{\text{EH}}^H) \ge \Gamma_{\text{EH}}, \ \mathrm{tr}(\boldsymbol{S}) \le P,
\end{equation}
whose feasibility requires that \(\boldsymbol{a}_t^T \boldsymbol{S} \boldsymbol{a}_t^\dag \ge \frac{1}{\|\boldsymbol{\dot{a}}_r\|^2 \Gamma_{\text{S,1}}}\) holds. On the Pareto boundary, the C-min and R-max vertices coincide, as both are found by maximizing \(\boldsymbol{a}_t^T \boldsymbol{S} \boldsymbol{a}_t^\dag\). Also, the C-E and R-E tradeoff edges coincide, since they can be acquired by solving \eqref{31}.

\subsubsection{Co-located C-E setup}

In this setup, we perform SWIPT and locate the EH receiver concurrently. Accordingly, the LoS WPT channel is \(\boldsymbol{H}_{\text{EH}} = \alpha_{\text{EH}} \boldsymbol{a}_{\text{EH}}(\theta) \boldsymbol{a}_t^T(\theta)\), where \(\alpha_{\text{EH}}\) is the path loss and \(\boldsymbol{a}_{\text{EH}}(\theta) = \Big[e^{-j \frac{N_{\text{EH}}-1}{2} \pi \sin \theta}, e^{-j \frac{N_{\text{EH}}-3}{2} \pi \sin \theta}, \dots, e^{j \frac{N_{\text{EH}}-1}{2} \pi \sin \theta}\Big]^T\) is the steering vector at the EH receiver.
Since \(\boldsymbol{a}_{\text{EH}}^H \boldsymbol{a}_{\text{EH}} = N_{\text{EH}}\), the harvested energy is given by \(E(\boldsymbol{S}) = \alpha_{\text{EH}}^2 N_{\text{EH}} \boldsymbol{a}_t^T \boldsymbol{S} \boldsymbol{a}_t^\dag\). We consider a special case with a single antenna equipped at the ID receiver, i.e., \(N_{\text{ID}} = 1\), and denote the ID channel vector as \(\boldsymbol{h}_{\text{ID}} \in \mathbb{C}^{M \times 1}\). In this case, similar to \eqref{31}, (P1) is simplified to
\begin{equation} \label{33}
	\begin{aligned}
		\max_{\boldsymbol{S} \succeq \boldsymbol{0}} &\ \boldsymbol{h}_{\text{ID}}^H \boldsymbol{S} \boldsymbol{h}_{\text{ID}} \\ 
		\mathrm{s.t.} &\ \boldsymbol{a}_t^T \boldsymbol{S} \boldsymbol{a}_t^\dag \ge \max\Big(\frac{\Gamma_{\text{EH}}}{\alpha_{\text{EH}}^2 N_{\text{EH}}}, \frac{1}{\|\boldsymbol{\dot{a}}_r\|^2 \Gamma_{\text{S,1}}}\Big), \ \mathrm{tr}(\boldsymbol{S}) \le P.
	\end{aligned}
\end{equation}
On the Pareto boundary, the C-min and E-max vertices coincide, and so do the C-R and R-E tradeoff edges, as they can be obtained by solving \eqref{33}.

\begin{figure}[tb]
	\centering 
	\subfloat[\footnotesize C-R-E view.] {\includegraphics[width=0.32\columnwidth]{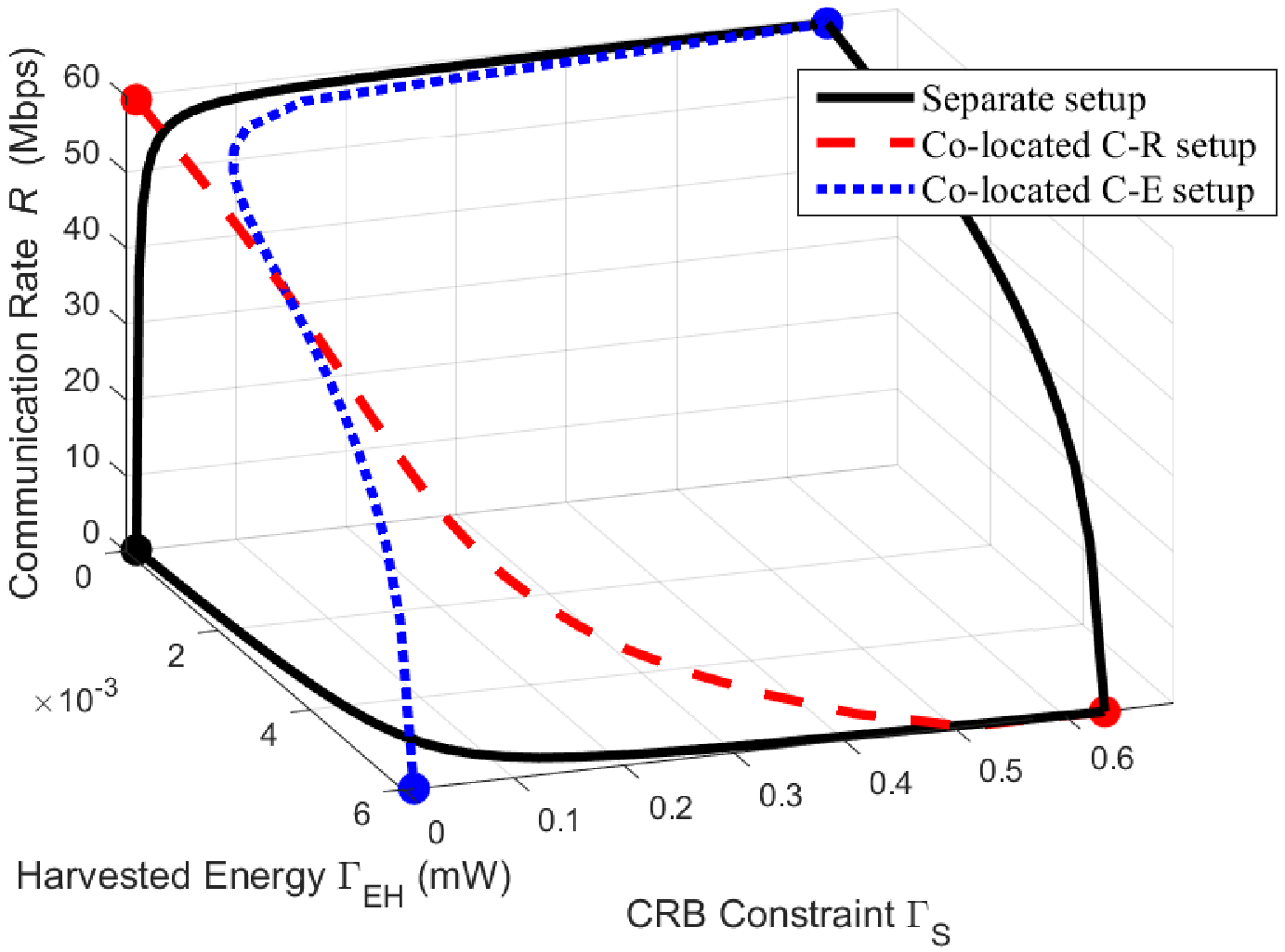}} 
	\subfloat[\footnotesize C-R view.] {\includegraphics[width=0.32\columnwidth]{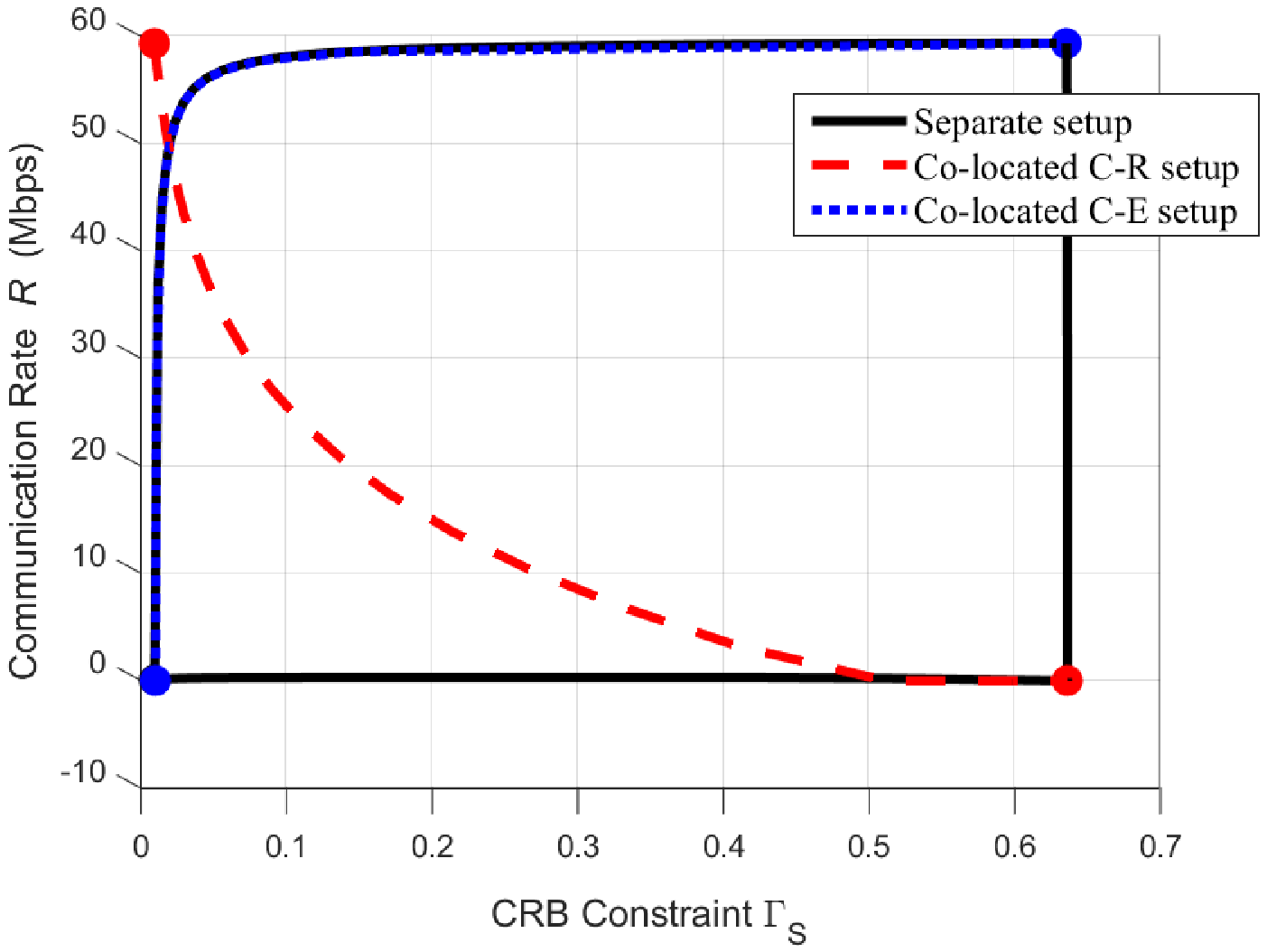}} 
	\subfloat[\footnotesize C-E view.] {\includegraphics[width=0.32\columnwidth]{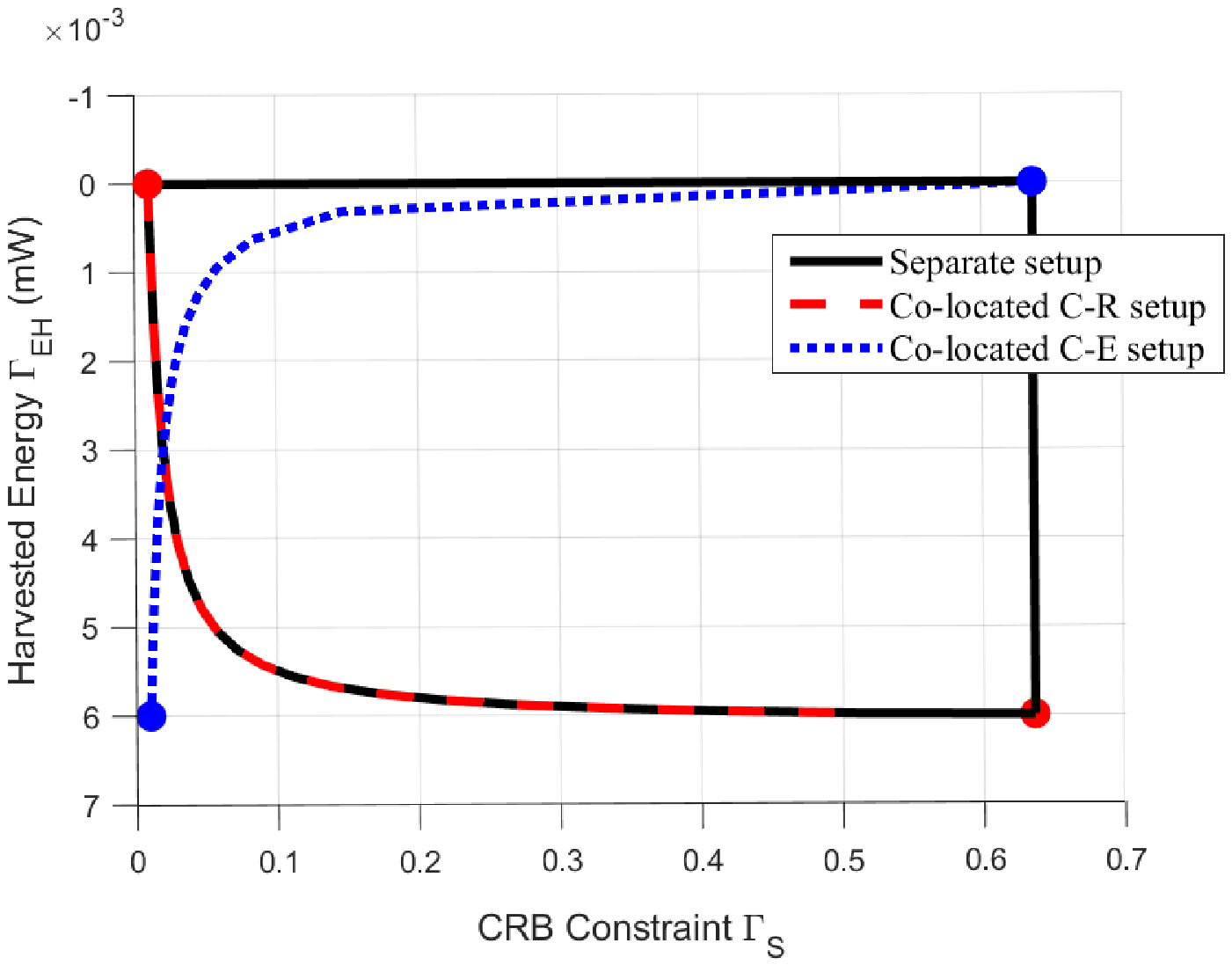}}
	\caption{The Pareto boundaries of the C-R-E regions for the separate setup and co-located C-R and C-E setups.} 
\end{figure}

The following simulation verifies the observations above. For the simulations in this paper, the H-AP is equipped with a ULA of \(M = 6\) and \(N_{\text{S}} = 16\) antennas with half a wavelength spacing between adjacent antennas. The target reflection coefficient is set as \(\alpha = 10^{-6}\), accounting for a round-trip path loss of \(120\) dB. The transmit power at the H-AP is \(30\) dBm. The noise powers at the sensing receiver and the ID receiver over the transmitted signal bandwidth of \(100\) KHz are set to be \(\sigma_{\text{S}}^2 = \sigma_{\text{ID}}^2 = -90\) dBm. For the EH receiver, the energy conversion efficiency is \(\zeta = 50 \)\% \cite{zhang2013mimo}. The transmission duration is set as \(L = 256\).

Fig. 3 shows the Pareto boundaries of the C-R-E regions for the point target case with \(N_{\text{ID}} = 1\) and \(N_{\text{EH}} = 2\). We consider the line-of-sight (LoS) ID and EH channels with \(\alpha_{\text{ID}} = \alpha_{\text{EH}} = 10^{-3}\). For comparison, for the separate setup, we set \(\sin \theta = \frac{1}{6}\), \(\sin \theta_{\text{ID}} = \frac{5}{6}\), and \(\sin \theta_{\text{EH}} = \frac{3}{2}\). For the co-located C-R and C-E setups, we set \(\theta_{\text{ID}} = \theta\) and \(\theta_{\text{EH}} = \theta\), respectively, such that the Pareto boundary surfaces collapse into curves. To be specific, with co-located C-R, the original C-R-E tradeoff boils down to the C\&R-E tradeoff,\footnote{By decomposing \(\boldsymbol{S} = \boldsymbol{s} \boldsymbol{s}^H\), problem \eqref{31} can be reexpressed as \(\max_{\boldsymbol{s}} |\boldsymbol{a}_t^T \boldsymbol{s}|^2, \ \mathrm{s.t.} \ \sum_{i=1}^{N_{\text{EH}}} \lambda_{\text{EH},i} |\boldsymbol{v}_{\text{EH},i}^H \boldsymbol{s}|^2 \ge \Gamma_{\text{EH}}, \ \|\boldsymbol{s}\|^2 \le P\), where \(\{\lambda_{\text{EH},i}\}\) and \(\{\boldsymbol{v}_{\text{EH},i}\}\) are the eigenvalues and corresponding eigenvectors of \(\boldsymbol{H}_{\text{EH}}^H \boldsymbol{H}_\text{EH}\), respectively. In this case, the optimal solution satisfies \(\boldsymbol{s} \in \mathrm{span}\big\{\boldsymbol{a}_t^\dag,\{\boldsymbol{v}_{\text{EH},i}\}\big\}\).} which enables improved ISAC performance than the separate setup under the same harvested energy; with co-located C-E, the C-R-E tradeoff is merged into C\&E-R tradeoff,\footnote{Similar to \eqref{31}, \eqref{33} can be reexpressed as \(\max_{\boldsymbol{s}} |\boldsymbol{h}_{\text{ID}}^H \boldsymbol{s}|^2, \ \mathrm{s.t.} \ |\boldsymbol{a}_t^T \boldsymbol{s}|^2 \ge \max(\frac{\Gamma_{\text{EH}}}{\alpha_{\text{EH}}^2 N_{\text{EH}}}, \frac{1}{\|\boldsymbol{\dot{a}}_r\|^2 \Gamma_{\text{S,1}}}), \ \|\boldsymbol{s}\|^2 \le P\). The closed-form optimal solution has been given in \cite{liu2021cramer}, which satisfies \(\boldsymbol{s} \in \mathrm{span}\{\boldsymbol{a}_t^\dag,\boldsymbol{h}_{\text{ID}}\}\).} which enhances sensing and WPT performance compared to the separate setup at a given communication rate.

\section{Optimal Solution to Problem (P2) with Extended Target}

In this section, we address problem (P2) for the extended target case. For convenience, we rewrite the constraint in \eqref{Rb} for \(i = 2\) as 
\begin{equation} \label{RRb-reformulation}
	\mathrm{tr}(\boldsymbol{S}^{-1}) \le \Gamma_{\text{S},2},
\end{equation}
where \(\Gamma_{\text{S},2} \triangleq \frac{L}{\sigma_{\text{S}}^2 N_{\text{S}}} \Gamma_{\text{S}}\). Notice that (P2) is convex and thus can be solved by using standard convex optimization techniques \cite{grant2014cvx}. In the following, we develop an efficient algorithm to obtain a partitionable solution of \(\boldsymbol{S}\) to (P2).

\subsection{Optimal Solution to (P2) Based on Problem Reformulation}

This subsection finds the optimal solution to problem (P2) based on its reformulation. We define \(\boldsymbol{H} \triangleq \big[\boldsymbol{H}_{\text{ID}}^T \ \boldsymbol{H}_{\text{EH}}^T\big]^T \in \mathbb{C}^{(N_{\text{ID}}+N_{\text{EH}}) \times M}\), which corresponds to the combined ID and EH channel matrix with \(\mathrm{rank}(\boldsymbol{H}) = r\). The SVD of \(\boldsymbol{H}\) is expressed as
\begin{equation} \label{svdH}
	\boldsymbol{H} = 
	\begin{bmatrix}
		\boldsymbol{U}^{\overline{\mathrm{null}}} \ \boldsymbol{U}^{\mathrm{null}}
	\end{bmatrix}
	\begin{bmatrix}
		\boldsymbol{\Lambda}^{\overline{\mathrm{null}}} & \boldsymbol{0} \\ \boldsymbol{0} & \boldsymbol{0}
	\end{bmatrix}
	\begin{bmatrix}
		{\boldsymbol{V}^{\overline{\mathrm{null}}}}^H \\ {\boldsymbol{V}^{\mathrm{null}}}^H
	\end{bmatrix},
\end{equation}
where \(\boldsymbol{\Lambda}^{\overline{\mathrm{null}}} = \mathrm{diag}(\lambda_1, \dots, \lambda_{r})\), and \({\boldsymbol{V}^{\overline{\mathrm{null}}}} \in \mathbb{C}^{M \times r}\) and \({\boldsymbol{V}^{\mathrm{null}}} \in \mathbb{C}^{M \times (M-r)}\) denote the right singular vectors corresponding to the \(r\) non-zero singular values and the \(M-r\) zero singular values of \(\boldsymbol{H}\), respectively. Without loss of generality, we express \(\boldsymbol{S}\) as
\begin{equation} \label{S_e}
		\boldsymbol{S} = \begin{bmatrix}
			\boldsymbol{V}^{\overline{\mathrm{null}}} \ \boldsymbol{V}^{\mathrm{null}}
		\end{bmatrix}
		\begin{bmatrix}
			\boldsymbol{S}_1 & \boldsymbol{B} \\ \boldsymbol{B}^H & \boldsymbol{S}_0
		\end{bmatrix}
		\begin{bmatrix}
			{\boldsymbol{V}^{\overline{\mathrm{null}}}}^H \\ {\boldsymbol{V}^{\mathrm{null}}}^H
		\end{bmatrix},
\end{equation}
where \(\boldsymbol{S}_1 \in \mathbb{S}^r\), \(\boldsymbol{S}_0 \in \mathbb{S}^{M-r}\), and \(\boldsymbol{B} \in \mathbb{C}^{r \times (M-r)}\) are variables to be optimized. Note that since \(\boldsymbol{H} \boldsymbol{V}^{\mathrm{null}} = \boldsymbol{0}\), problem (P2) is equivalent to
\begin{subequations}
	\begin{align}
		(\text{P}2.1): \max_{\boldsymbol{S}_1, \boldsymbol{S}_0, \boldsymbol{B}} &\ \log_2 \det \Big(\boldsymbol{I}_{N_{\text{ID}}} + \frac{\boldsymbol{H}_{\text{ID}} \boldsymbol{V}^{\overline{\mathrm{null}}} \boldsymbol{S}_1 {\boldsymbol{V}^{\overline{\mathrm{null}}}}^H \boldsymbol{H}_{\text{ID}}^H}{\sigma_{\text{ID}}^2}\Big) \label{RRR}\\
		\mathrm{s.t.} &\ \mathrm{tr}(\boldsymbol{H}_{\text{EH}} \boldsymbol{V}^{\overline{\mathrm{null}}} \boldsymbol{S}_1 {\boldsymbol{V}^{\overline{\mathrm{null}}}}^H \boldsymbol{H}_{\text{EH}}^H) \ge \Gamma_{\text{EH}}, \label{RRRa} \\
		&\ \mathrm{tr}\left(\begin{bmatrix}
			\boldsymbol{S}_1 & \boldsymbol{B} \\ \boldsymbol{B}^H & \boldsymbol{S}_0
		\end{bmatrix}^{-1}\right) \le \Gamma_{\text{S},2}, \label{RRRb}\\
		&\ \mathrm{tr}(\boldsymbol{S}_1) + \mathrm{tr}(\boldsymbol{S}_0) \le P, \label{RRRc}\\
		&\ \begin{bmatrix}
			\boldsymbol{S}_1 & \boldsymbol{B} \\ \boldsymbol{B}^H & \boldsymbol{S}_0
		\end{bmatrix} \succeq \boldsymbol{0}. \label{RRRd}
	\end{align}
\end{subequations}
To obtain a more insightful solution, we eliminate \(\boldsymbol{B}\) and rewrite \(\boldsymbol{S}_0\) in constraints \eqref{RRRb}, \eqref{RRRc}, and \eqref{RRRd} as follows.

\begin{proposition} \label{prop3}
	Problem (P2.1) is equivalent to the following problem.
	\begin{equation*}
		\begin{aligned}
			(\text{P}2.2): \max_{\boldsymbol{S}_1 \succeq \boldsymbol{0}, p_{0} \ge 0} &\ \log_2 \det \Big(\boldsymbol{I}_{N_{\text{ID}}} + \frac{1}{\sigma_{\text{ID}}^2} \widetilde{\boldsymbol{H}}_{\text{ID}} \boldsymbol{S}_1 \widetilde{\boldsymbol{H}}_{\text{ID}}^H\Big) \\
			\mathrm{s.t.} &\ \mathrm{tr}(\widetilde{\boldsymbol{H}}_{\text{EH}} \boldsymbol{S}_1 \widetilde{\boldsymbol{H}}_{\text{EH}}^H) \ge \Gamma_{\text{EH}}, \\
			&\ \mathrm{tr}(\boldsymbol{S}_1^{-1}) + \frac{M-r}{p_{0}} \le \Gamma_{\text{S},2}, \\
			&\ \mathrm{tr}(\boldsymbol{S}_1) + (M-r) p_{0} \le P,
		\end{aligned}
	\end{equation*}
	where \(\widetilde{\boldsymbol{H}}_{\text{ID}} \triangleq \boldsymbol{H}_{\text{ID}} \boldsymbol{V}^{\overline{\mathrm{null}}}\) and \(\widetilde{\boldsymbol{H}}_{\text{EH}} \triangleq \boldsymbol{H}_{\text{EH}} \boldsymbol{V}^{\overline{\mathrm{null}}}\) denotes the projection of the ID and EH channels on the range space of the combined channel \(\boldsymbol{H}\), respectively.
\end{proposition}

\textit{Proof:} See Appendix B.
\hfill \(\square\)

Let \(\boldsymbol{S}_1^{\mathrm{opt}}\) and \(p_0^{\mathrm{opt}}\) denote the optimal solution to (P2.2). Then the optimal solution to (P2) is
\begin{equation} \label{S_opt2}
	\boldsymbol{S}^{\mathrm{opt},2} = \boldsymbol{V}^{\overline{\mathrm{null}}} \boldsymbol{S}_1^{\mathrm{opt}} {\boldsymbol{V}^{\overline{\mathrm{null}}}}^H + p_0^{\mathrm{opt}} \boldsymbol{V}^{\mathrm{null}} {\boldsymbol{V}^{\mathrm{null}}}^H.
\end{equation}

\begin{remark} \label{rem2}
	Based on \eqref{S_opt2}, the optimal transmit covariance matrix \(\boldsymbol{S}^{\mathrm{opt},2}\) for the extended target case differs from that of \(\boldsymbol{S}^{\mathrm{opt},1}\) to problem (P1) for the point target case in several aspects. Firstly, while \(\boldsymbol{S}^{\mathrm{opt},1}\) is derived based on SVD of the ID channel \(\boldsymbol{H}_{\text{ID}}\) and the composite sensing and EH channel \(\boldsymbol{D}\), \(\boldsymbol{S}^{\mathrm{opt},2}\) is obtained based on SVD of the combined ID and EH channel \(\boldsymbol{H} = \big[\boldsymbol{H}_{\text{ID}}^T \ \boldsymbol{H}_{\text{EH}}^T\big]^T\). Consequently, \(\boldsymbol{S}^{\mathrm{opt},2}\) is divided into two parts. The first part, \(\boldsymbol{V}^{\overline{\mathrm{null}}} \boldsymbol{S}_1^{\mathrm{opt}} {\boldsymbol{V}^{\overline{\mathrm{null}}}}^H\), balances the tradeoff among the WF power allocation for communication, the strongest EMT for WPT, and the isotropic transmission for sensing. This part requires \(\boldsymbol{S}_1^{\mathrm{opt}}\) to be a general positive semidefinite matrix, which differs from the diagonal structure of \(\boldsymbol{S}_{11}^{\mathrm{opt}}\) to (P1). The second part, \(p_0^{\mathrm{opt}} \boldsymbol{V}^{\mathrm{null}} {\boldsymbol{V}^{\mathrm{null}}}^H\), is dedicated to sensing only, which differs from \(\boldsymbol{C}^{\mathrm{opt}}\) and \(\boldsymbol{S}_{00}^{\mathrm{opt}}\) to (P1) for both sensing and WPT. Furthermore, the CRB constraint in \eqref{RRb-reformulation} requires \(\boldsymbol{S}^{\mathrm{opt},2}\) to be full-rank and thus the average transmit power allocated for dedicated sensing \(p_0^{\mathrm{opt}}\) is always necessary for cases with \(M>r\), whereas \(\boldsymbol{C}^{\mathrm{opt}}\) and \(\boldsymbol{S}_{00}^{\mathrm{opt}}\) to (P1) are generally not required for randomly generated channels.
\end{remark}

\subsection{Semi-Closed-Form Solution to Problem (P2) in Special Case with Low SNR}

To obtain more insights, this subsection further considers the special case with low communication SNR (i.e., \(\frac{P}{\sigma_{\text{ID}}^2} \mathrm{tr}(\boldsymbol{H}_{\text{ID}}^H \boldsymbol{H}_{\text{ID}}) \to 0\)) and obtains a structured solution to problem (P2) in this case. Towards this end, we transform problem (P2) as problem (P2.2) similarly as in Section V-A, and then focus on solving (P2.2). 
In (P2.2), maximizing \(\log_2 \det \Big(\boldsymbol{I}_{N_{\text{ID}}} + \frac{1}{\sigma_{\text{ID}}^2} \widetilde{\boldsymbol{H}}_{\text{ID}} \boldsymbol{S}_1 \widetilde{\boldsymbol{H}}_{\text{ID}}^H\Big)\) is equivalent to maximizing \(\mathrm{tr}(\widetilde{\boldsymbol{H}}_{\text{ID}} \boldsymbol{S}_1 \widetilde{\boldsymbol{H}}_{\text{ID}}^H)\) due to the consideration of low SNR. Accordingly, (P2.2) is reformulated as
\begin{subequations}
	\begin{align}
		(\text{P}3): \max_{\boldsymbol{S}_1 \succeq \boldsymbol{0}, p_{0} \ge 0} &\ \mathrm{tr}(\widetilde{\boldsymbol{H}}_{\text{ID}} \boldsymbol{S}_1 \widetilde{\boldsymbol{H}}_{\text{ID}}^H) \label{RRRRR} \\
		\mathrm{s.t.} &\ \mathrm{tr}(\widetilde{\boldsymbol{H}}_{\text{EH}} \boldsymbol{S}_1 \widetilde{\boldsymbol{H}}_{\text{EH}}^H) \ge \Gamma_{\text{EH}} \label{RRRRRa}, \\
		&\ \mathrm{tr}(\boldsymbol{S}_1^{-1}) + (M-r) \frac{1}{p_{0}} \le \Gamma_{\text{S},2} \label{RRRRRb}, \\
		&\ \mathrm{tr}(\boldsymbol{S}_1) + (M-r) p_{0} \le P. \label{RRRRRc}
	\end{align}
\end{subequations}

\begin{proposition} \label{prop5}
	The solution to problem (P3) is
	\begin{equation} \label{Slow}
		\boldsymbol{S}_1^{\mathrm{low}} = \boldsymbol{Q}_{\text{e}} \mathrm{diag}\Big(\sqrt{\frac{\mu^*}{\sigma_{\text{e},1}}}, \dots, \sqrt{\frac{\mu^*}{\sigma_{\text{e},r}}}\Big) {\boldsymbol{Q}_{\text{e}}}^H, \
		p_{0}^{\mathrm{low}} = \sqrt{\frac{\mu^*}{\nu^*}},
	\end{equation}
	where \(\lambda^* \ge 0\), \(\mu^* > 0\), and \(\nu^* > 0\) denote the optimal dual variables associated with the constraints in \eqref{RRRRRa}, \eqref{RRRRRb}, and \eqref{RRRRRc} in (P3), respectively, and \({\boldsymbol{Q}_{\text{e}}}\) and \(\{\sigma_{\text{e},k}\}, \forall k \in \{1, \dots, r\}\), are obtained based on the EVD \(\nu^* \boldsymbol{I}_{r} - \widetilde{\boldsymbol{H}}_{\text{ID}}^H \widetilde{\boldsymbol{H}}_{\text{ID}} - \lambda^* \widetilde{\boldsymbol{H}}_{\text{EH}}^H \widetilde{\boldsymbol{H}}_{\text{EH}} = \boldsymbol{Q}_{\text{e}} \mathrm{diag}(\sigma_{\text{e},1}, \dots, \sigma_{\text{e},r}) {\boldsymbol{Q}_{\text{e}}}^H\).
\end{proposition}

\textit{Proof:} See Appendix C.
\hfill \(\square\)

Based on Proposition \ref{prop5} and \eqref{S_opt2}, we obtain the optimal transmit covariance matrix in the low SNR case as \(\boldsymbol{S}^{\mathrm{low}} = \boldsymbol{V}^{\overline{\mathrm{null}}} \boldsymbol{S}_1^{\mathrm{low}} {\boldsymbol{V}^{\overline{\mathrm{null}}}}^H + p_0^{\mathrm{low}} \boldsymbol{V}^{\mathrm{null}} {\boldsymbol{V}^{\mathrm{null}}}^H\), which is also of full rank to satisfy the CRB requirement. It is observed in \eqref{Slow} that \(\boldsymbol{S}_1^{\mathrm{low}}\) follows the EMT structure based on the composite channel matrix \(\nu^* \boldsymbol{I}_{r} - \widetilde{\boldsymbol{H}}_{\text{ID}}^H \widetilde{\boldsymbol{H}}_{\text{ID}} - \lambda^* \widetilde{\boldsymbol{H}}_{\text{EH}}^H \widetilde{\boldsymbol{H}}_{\text{EH}}\), together with a channel-inversion-like power allocation, to balance the competing between the isotropic transmission for sensing and the strongest EMT for combined communication and WPT (or SWIPT), with \(\{\sigma_{\text{e},k}\}, \forall k \in \{1, \dots, r\}\), denoting the equivalent channel power gains.

\subsection{Special Case with Co-located R-E}
In the co-located R-E setup, we perform SWIPT for a co-located ID\&EH receiver and locate a separate target concurrently. Accordingly, the communication and WPT channels are identical, i.e., \(\boldsymbol{H}_{\text{ID}} = \boldsymbol{H}_{\text{EH}} = \boldsymbol{G}\) and \(N_{\text{ID}} = N_{\text{EH}} = N\). In the following, we investigate two co-located receiver designs with time-switching and power-splitting.

\subsubsection{Time-switching design} 

For this type of receiver, we define the ratio of time allocated for ID and EH as \(1-\tau\) and \(\tau\), respectively, where \(0<\tau<1\). We also define the transmit covariance matrices for ID and EH as \(\boldsymbol{\bar{S}}_1\) and \(\boldsymbol{\bar{S}}_2\), respectively. With the flexible power constraint, the SVD of \(\boldsymbol{G} = \boldsymbol{U} \boldsymbol{\Lambda} \boldsymbol{V}^H\), and defining \(\boldsymbol{\widetilde{S}}_1 = \boldsymbol{V} \boldsymbol{\bar{S}}_1  \boldsymbol{V}^H\) and \(\boldsymbol{\widetilde{S}}_2 = \boldsymbol{V} \boldsymbol{\bar{S}}_2  \boldsymbol{V}^H\), (P2) becomes \(\max_{\boldsymbol{\widetilde{S}}_1 \succeq \boldsymbol{0}, \boldsymbol{\widetilde{S}}_2 \succeq \boldsymbol{0}} \ (1-\tau) \log_2 \det \Big(\boldsymbol{I}_{N} + \frac{1}{\sigma_{\text{ID}}^2} \boldsymbol{\Lambda}^H \boldsymbol{\Lambda} \boldsymbol{\widetilde{S}}_1\Big)\), \(\mathrm{s.t.} \ \tau \mathrm{tr}(\boldsymbol{\Lambda}^H \boldsymbol{\Lambda} \boldsymbol{\widetilde{S}}_2) \ge \Gamma_{\text{EH}}, \ \mathrm{tr}\Big(\big((1-\tau) \boldsymbol{\widetilde{S}}_1 + \tau \boldsymbol{\widetilde{S}}_2\big)^{-1}\Big) \le \Gamma_{\text{S},2}, \ \mathrm{tr}((1-\tau) \boldsymbol{\widetilde{S}}_1 + \tau \boldsymbol{\widetilde{S}}_2) \le P\). According to Proposition 3, the optimal solutions to \(\boldsymbol{\widetilde{S}}_1\) and \(\boldsymbol{\widetilde{S}}_2\) admit the form of \(\boldsymbol{\widetilde{S}}_i = \mathrm{diag}(p_{i,1}, \dots, p_{i,\bar{r}}, p_{i,0}, \dots, p_{i,0})\), \(\forall i = 1,2\), where \(\bar{r} = \mathrm{rank}(\boldsymbol{G})\). In particular, if the receiver is equipped with a single antenna, i.e., \(\boldsymbol{G}\) becomes \(\boldsymbol{g}^H \in \mathbb{C}^{1 \times M}\), then the optimal solution admits the form of \(\boldsymbol{S}_i = p_{i,1} \frac{\boldsymbol{g} \boldsymbol{g}^H}{\|\boldsymbol{g}\|^2} + p_{i,0} (\boldsymbol{I}_M - \frac{\boldsymbol{g} \boldsymbol{g}^H}{\|\boldsymbol{g}\|^2}), \forall i = 1,2\), and (P2) is equivalent to \(\max_{p_{1,1},p_{1,0},p_{2,1},p_{2,0}} \ p_{1,1}, \ \mathrm{s.t.} \ \tau \|\boldsymbol{g}\|^2 p_{2,1} \ge \Gamma_{\text{EH}}, \ \frac{1}{(1-\tau) p_{1,1} + \tau p_{2,1}} + \frac{M-1}{(1-\tau) p_{1,0} + \tau p_{2,0}}\le \Gamma_{\text{S},2}, \ (1-\tau) p_{1,1} + \tau p_{2,1} + (M-1)\big((1-\tau) p_{1,0} + \tau p_{2,0}\big) \le P\). Applying the Lagrangian dual method, we find that all the three constraints are tight at the optimality, thus the Pareto boundary with given \(\tau\) is \(\big\{(\mathrm{CRB}_{\text{TS}}, R_{\text{TS}}, E_{\text{TS}}) \big| \mathrm{CRB}_{\text{TS}} = \frac{\sigma_{\text{S}}^2 N_{\text{S}} \Gamma_{\text{S},2}}{L}, R_{\text{TS}} = (1-\tau)\log_2(1+\frac{\Gamma_1}{(1-\tau) \sigma_{\text{ID}}^2}), E_{\text{TS}} = \Gamma_2, P = \frac{\Gamma_1+\Gamma_2}{\|\boldsymbol{g}\|^2} + \frac{(M-1)^2}{\Gamma_{\text{S},2} - \frac{\|\boldsymbol{g}\|^2}{\Gamma_1+\Gamma_2}}\big\}\).

\subsubsection{Power-splitting design}

For this type of receiver, we define the ratio of power allocated for ID and EH at each receive antenna \(n \in \{1, \dots, N\}\) as \(1-\rho_n\) and \(\rho_n\), respectively, where \(0<\rho_n<1\). Then, (P2) becomes \(\max_{\boldsymbol{S} \succeq \boldsymbol{0}} \ \log_2 \det \Big(\boldsymbol{I}_{N} + \frac{1}{\sigma_{\text{ID}}^2} \boldsymbol{\bar{\Delta}}^{1/2} \boldsymbol{G} \boldsymbol{S} \boldsymbol{G}^H \boldsymbol{\bar{\Delta}}^{1/2} \Big)\), \(\mathrm{s.t.} \ \mathrm{tr}(\boldsymbol{\Delta}^{1/2} \boldsymbol{G} \boldsymbol{S} \boldsymbol{G}^H \boldsymbol{\Delta}^{1/2}) \ge \Gamma_{\text{EH}}, \ \mathrm{tr}(\boldsymbol{S}^{-1}) \le \Gamma_{\text{S},2}, \ \mathrm{tr}(\boldsymbol{S}) \le P\), where \(\boldsymbol{\Delta} = \mathrm{diag}(\rho_1, \dots, \rho_N)\) and \(\boldsymbol{\bar{\Delta}} = \boldsymbol{I}_N - \boldsymbol{\Delta}\). By treating \(\boldsymbol{\bar{\Delta}}^{1/2} \boldsymbol{G}\) and \(\boldsymbol{\Delta}^{1/2} \boldsymbol{G}\) as equivalent ID and EH channels, this problem can be solved similarly as (P2) with given \(\rho_1, \dots, \rho_N\). If the receiver is equipped with a single antenna, then similar to the time-switching design, the optimal solution admits the form of \(\boldsymbol{S} = p_1 \frac{\boldsymbol{g} \boldsymbol{g}^H}{\|\boldsymbol{g}\|^2} + p_0 (\boldsymbol{I}_M - \frac{\boldsymbol{g} \boldsymbol{g}^H}{\|\boldsymbol{g}\|^2})\), and (P2) becomes \(\max_{p_1,p_0} \ p_1, \ \mathrm{s.t.} \ \frac{1}{p_1} + \frac{M-1}{p_0}\le \Gamma_{\text{S},2}, \ p_1 + (M-1) p_0 \le P\), whose feasibility requires \(\rho_1 \|\boldsymbol{g}\|^2 p_1 \ge \Gamma_{\text{EH}}\). According to Proposition 4, both the CRB and transmit power constraints are tight at the optimality, thus the Pareto boundary with given \(\rho_1\) is expressed as \(\big\{(\mathrm{CRB}_{\text{PS}}, R_{\text{PS}}, E_{\text{PS}}) \big| \mathrm{CRB}_{\text{PS}} = \frac{\sigma_{\text{S}}^2 N_{\text{S}} \Gamma_{\text{S},2}}{L}, R_{\text{PS}} = \log_2(1+\frac{(1-\rho_1) \Gamma}{\sigma_{\text{ID}}^2}),  E_{\text{PS}} = \rho_1 \Gamma, P = \frac{\Gamma}{\|\boldsymbol{g}\|^2} + \frac{(M-1)^2}{\Gamma_{\text{S},2} - \frac{\|\boldsymbol{g}\|^2}{\Gamma}}\big\}\).

\begin{figure}[tb]
	\centering 
	\subfloat[\footnotesize Cross-section view on \newline \(\Gamma_{\text{S}} = 20\mathrm{CRB}_{\mathrm{min},2}\).] {\includegraphics[width=0.45\columnwidth]{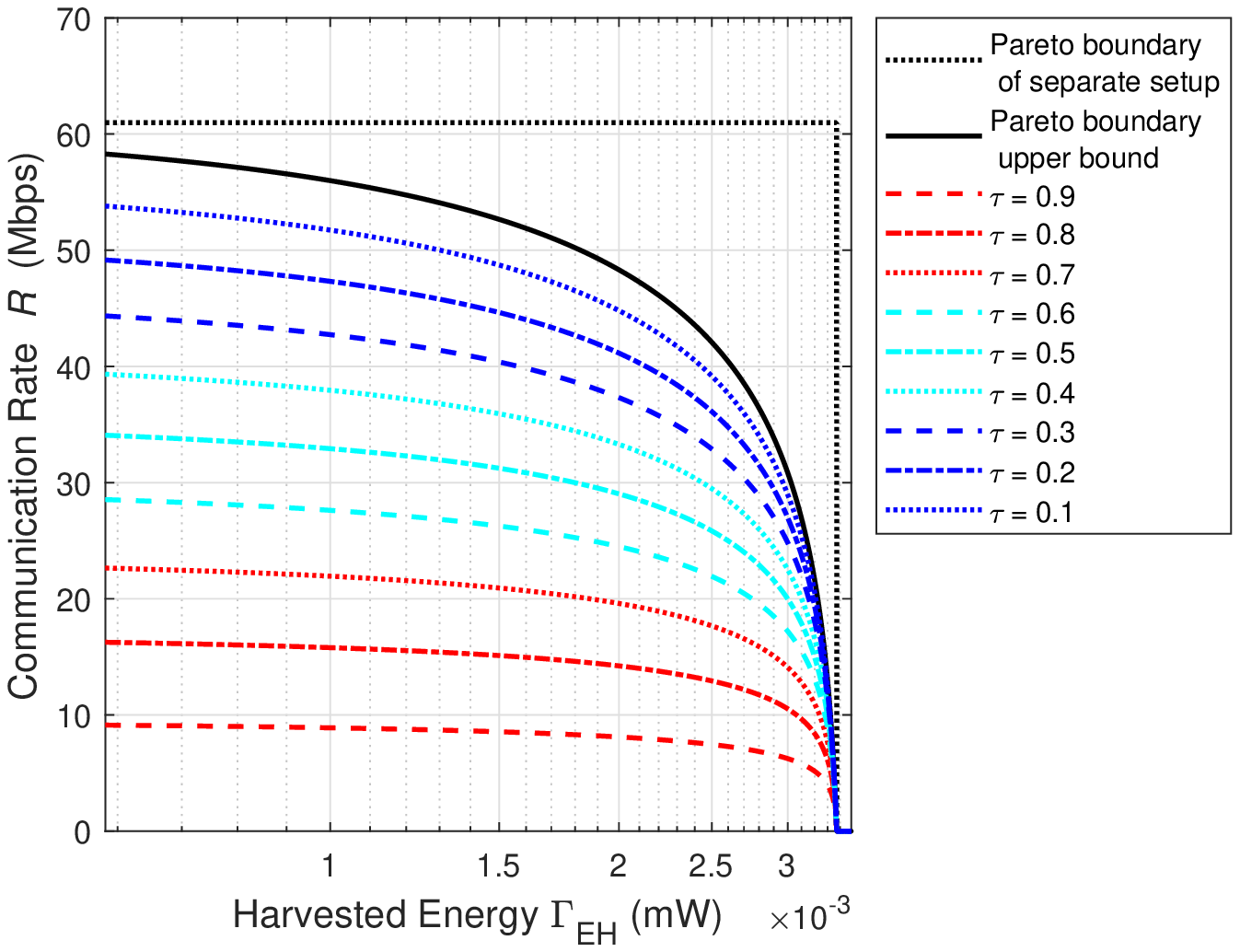}} 
	\subfloat[\footnotesize Cross-section view on \newline \(\Gamma_{\text{EH}} = 0.2E_{\mathrm{max}}\).] {\includegraphics[width=0.45\columnwidth]{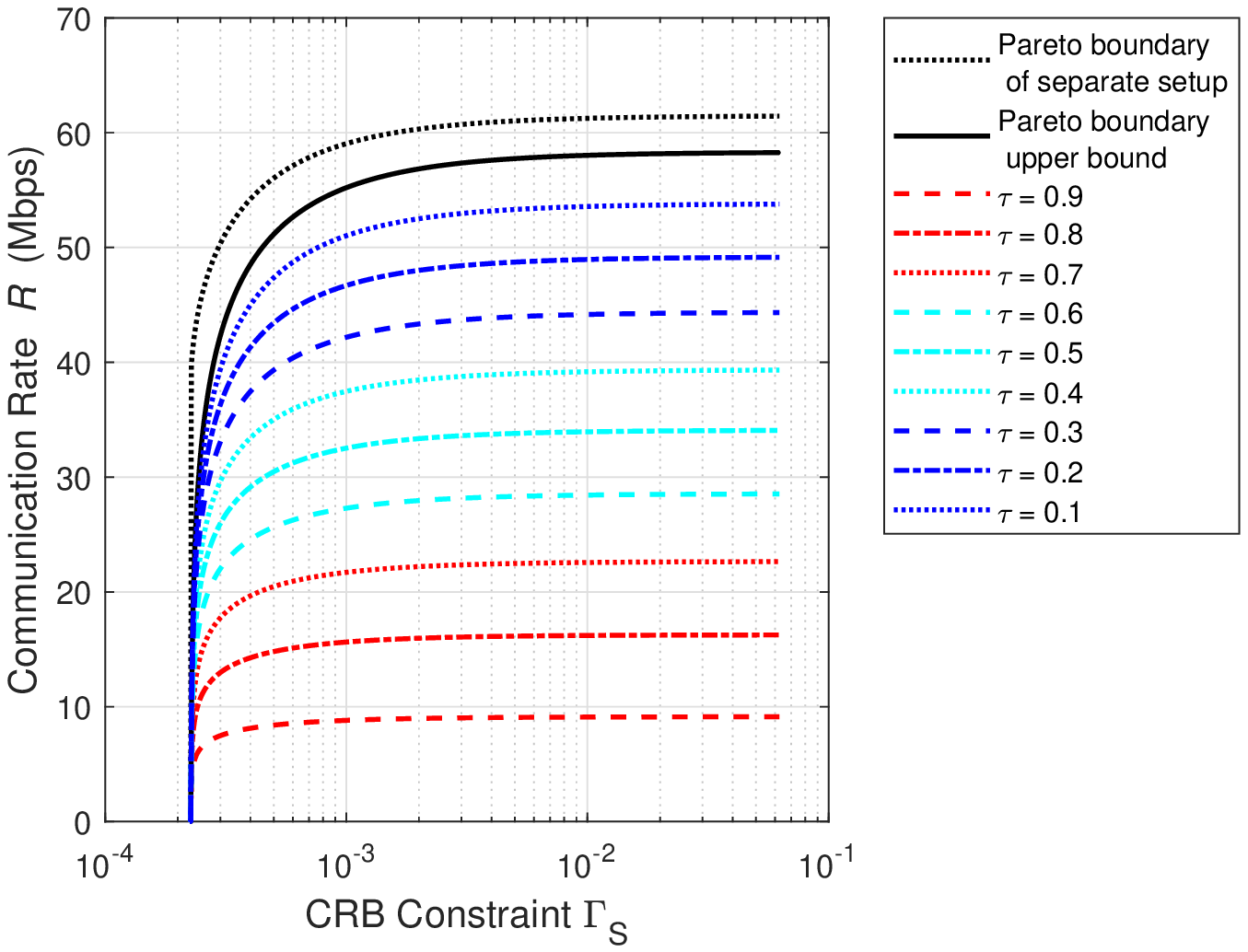}} 
	\caption{The Pareto boundary of the C-R-E region for the co-located R-E setup with the time-switching design.} 
\end{figure}

\begin{figure}[tb]
	\centering 
	\subfloat[\footnotesize Cross-section view on \newline \(\Gamma_{\text{S}} = 20\mathrm{CRB}_{\mathrm{min},2}\).] {\includegraphics[width=0.45\columnwidth]{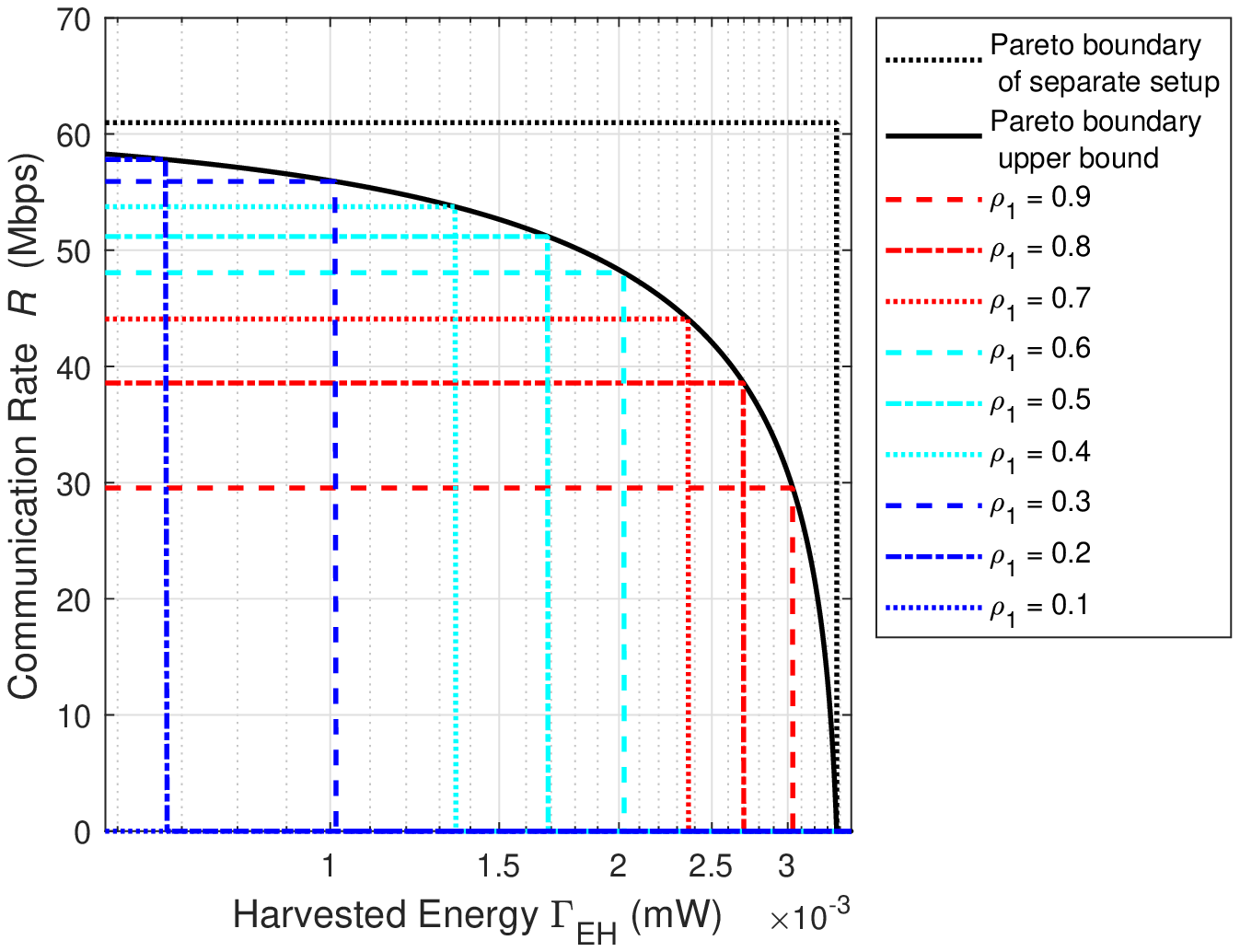}} 
	\subfloat[\footnotesize Cross-section view on \newline \(\Gamma_{\text{EH}} = 0.2E_{\mathrm{max}}\).] {\includegraphics[width=0.45\columnwidth]{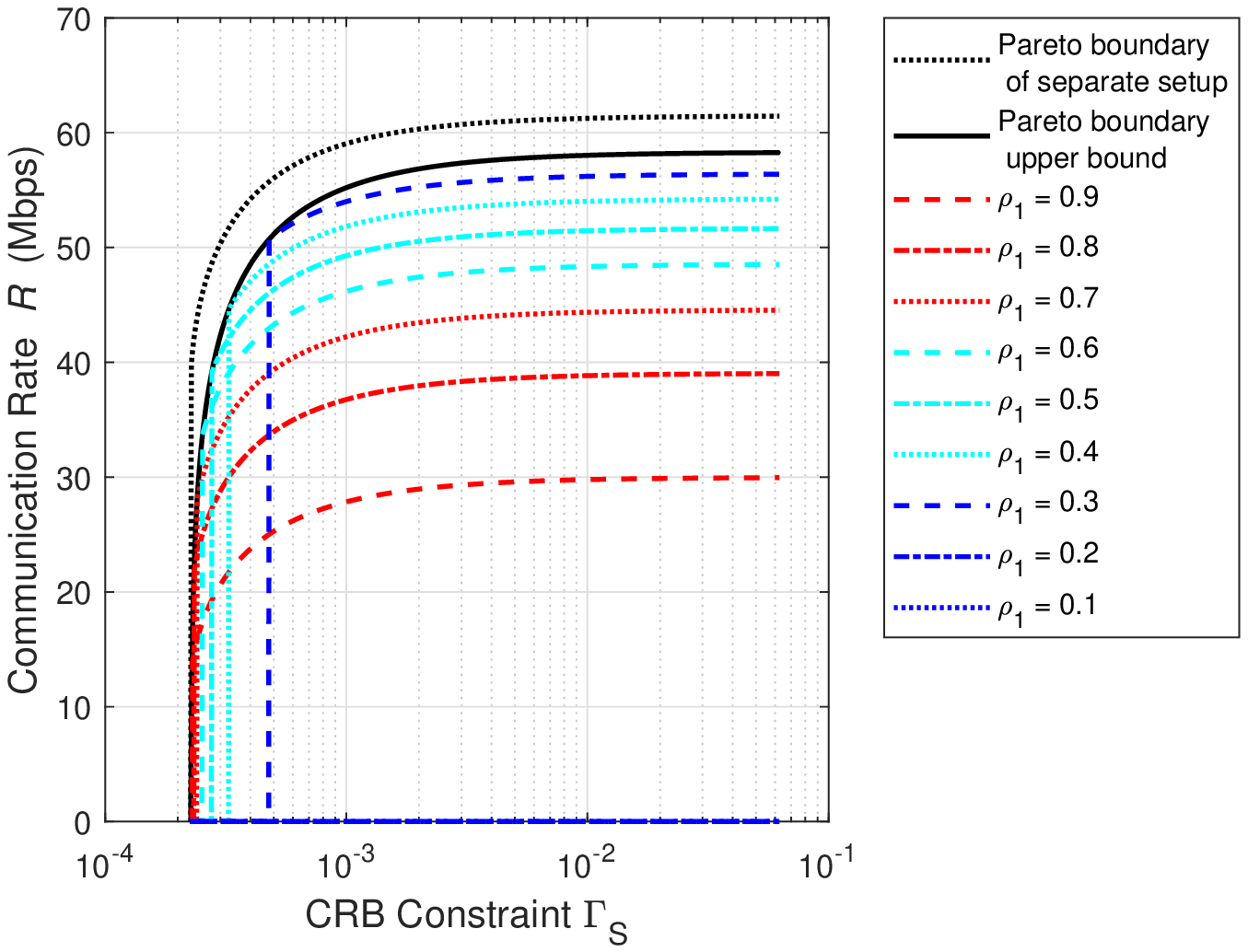}} 
	\caption{The Pareto boundary of the C-R-E region for the co-located R-E setup with the power-splitting design.} 
\end{figure}

Figs. 4 and 5 show the Pareto boundaries of the C-R-E regions for the extended target case with co-located R-E setup, where \(N_{\text{ID}} = N_{\text{EH}} = 1\). We consider the Rayleigh fading channel vector \(\boldsymbol{g}\), which is generated as a CSCG random vector with each element being zero mean and variance of \(10^{-6}\). As shown in Figs. 4 and 5,
the upper bound of the Pareto boundary for the time-switching design is achieved when \(\tau = 0\) (without the peak power constraint), which is the same as the union of Pareto boundaries for the power-splitting design with \(\rho_1\) ranging from 0 to 1.
We also show the Pareto boundary of the separate setup with \(\boldsymbol{g}\) modeling both ID and EH channels, which collapses to a curve, similar to the co-located C-R/C-E setups in Section IV-D. By contrast, for the co-located R-E setup, there is an R-E tradeoff at the receiver, which impairs the SWIPT performance. This tradeoff, along with the tradeoff between SWIPT and sensing, contributes to the gap between the Pareto boundary of the co-located R-E setup and that of the separate setup.

\section{Numerical Results}

This section evaluates the performance of our proposed optimal designs. In the following simulation, in addition to the system setups mentioned in Section IV-D, we consider \(\alpha_{\text{ID}} = 10^{-4}\) and \(\alpha_{\text{EH}} = 10^{-2}\) for the ID and EH channels, corresponding to the path loss of \(80\) dB and \(40\) dB \cite{zhang2013mimo}, respectively.

\begin{figure}[tb]
	\centering 
	{\includegraphics[width=0.30\textwidth]{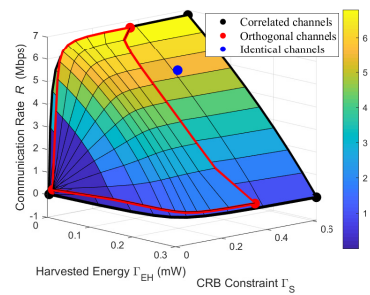}} 
	\caption{The Pareto boundaries of the C-R-E regions for the point target case with \(N_{\text{ID}} = N_{\text{EH}} = 1\).} 
\end{figure}
Fig. 6 shows the Pareto boundaries of the C-R-E regions for the point target case with \(N_{\text{ID}} = N_{\text{EH}} = 1\) and target angle \(\theta = 0\). In this figure, we consider the LoS channels for the ID and EH channels, i.e., \(\boldsymbol{H}_{\text{ID}} = \alpha_{\text{ID}} \boldsymbol{a}_t^T(\theta_{\text{ID}})\) and \(\boldsymbol{H}_{\text{EH}} = \alpha_{\text{EH}} \boldsymbol{a}_t^T(\theta_{\text{EH}})\), where \(\sin \theta_{\text{ID}} = \frac{2 \gamma}{M}\) and \(\sin \theta_{\text{EH}} = \frac{4 \gamma}{M}\) correspond to the angles of the ID and EH receivers, with $\gamma = 0$, $0.6$, and $1$ corresponding to the cases when the sensing, ID, and EH channels are identical, correlated, and orthogonal, respectively. It is observed that for the identical channels with $\gamma = 0$, the Pareto boundary collapses to a point, which means that the R-max, E-max, and C-min strategies are identical and the three performance metrics are optimized simultaneously. It is also observed that for the orthogonal channels with $\gamma = 1$, optimizing one metric (e.g., E-max) always leads to poor performances for the other two (e.g., a close-to-zero rate and the highest CRB), thus showing that the three objectives are conflicting with each other. Furthermore, for the correlated channels with $\gamma = 0.6$, the C-R-E region boundary is observed to lie between those with $\gamma = 1$ and $\gamma = 0$, thus showing the non-trivial C-R-E tradeoff in this case. 

\begin{figure}[tb]
	\centering 
	\subfloat[\footnotesize \(M = 6\).] {\includegraphics[width=0.45\columnwidth]{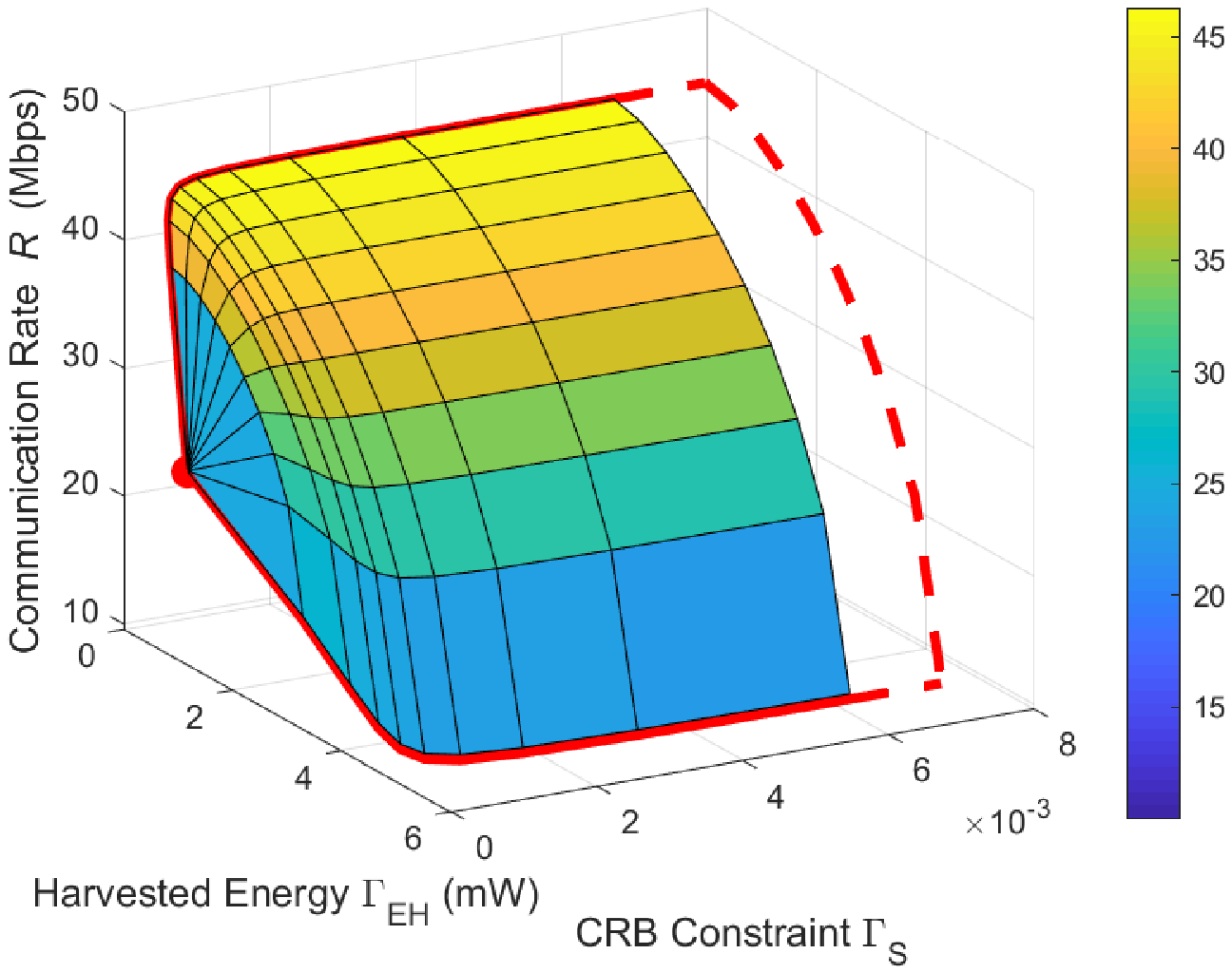}} 
	\subfloat[\footnotesize \(M = 2\).] {\includegraphics[width=0.45\columnwidth]{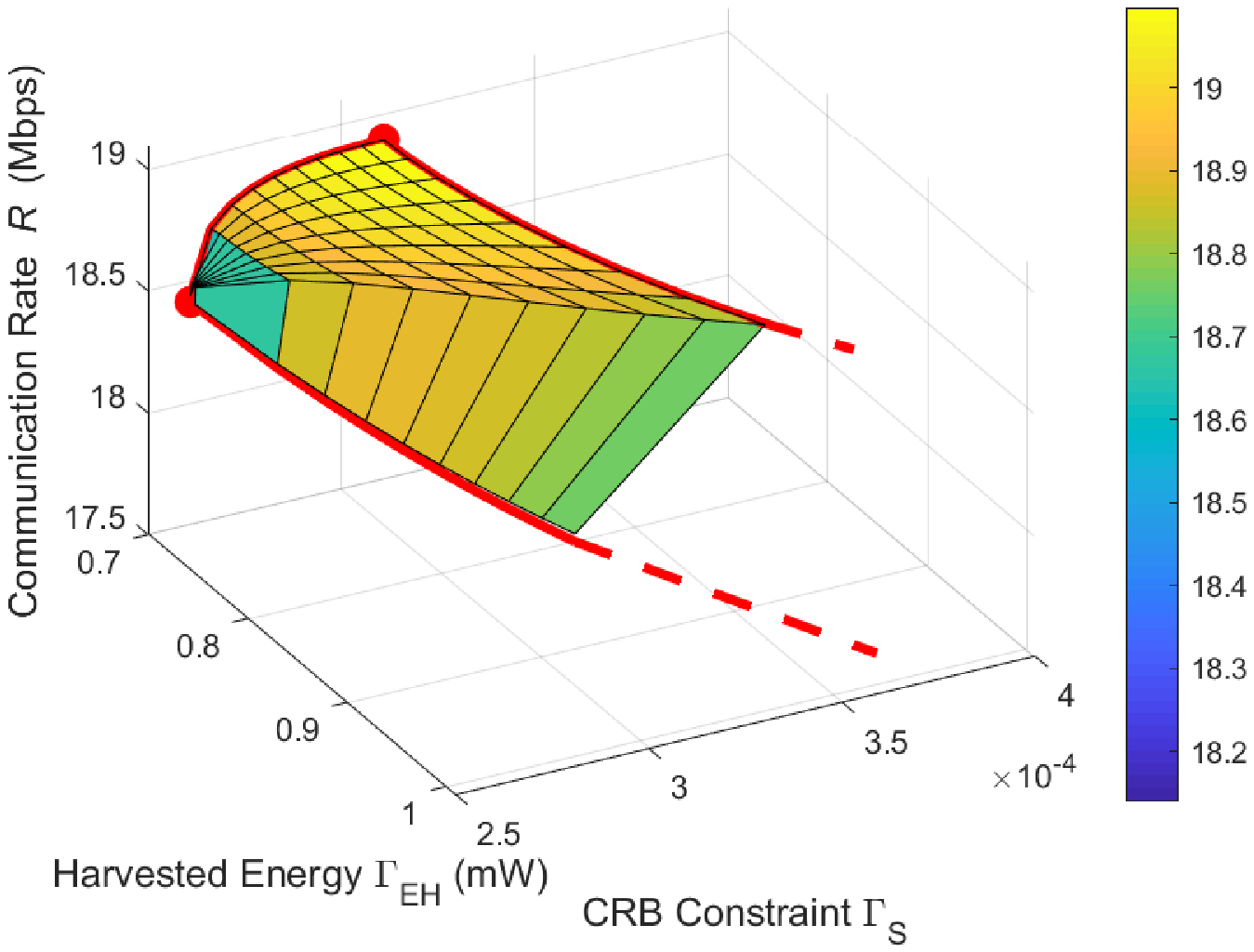}} 
	\caption{The Pareto boundaries of the C-R-E regions for the extended target case with \(N_{\text{ID}} = N_{\text{EH}} = 2\) and \(P = 40\) dBm.} 
\end{figure}
Figs. 7(a) and 7(b) show the Pareto boundaries of the C-R-E regions for the extended target case with \(M = 6\) and \(M = 2\), respectively, where \(P = 40\) dBm and \(N_{\text{ID}} = N_{\text{EH}} = 2\). We consider Rician fading for the ID and EH channels, i.e., \(\boldsymbol{H}_{\text{ID}} = \sqrt{\frac{\kappa_{\text{ID}}}{\kappa_{\text{ID}} + 1}} \boldsymbol{H}_{\text{ID}}^{\text{los}} + \sqrt{\frac{1}{\kappa_{\text{ID}} + 1}} \boldsymbol{H}_{\text{ID}}^w\) and \(\boldsymbol{H}_{\text{EH}} = \sqrt{\frac{\kappa_{\text{EH}}}{\kappa_{\text{EH}} + 1}} \boldsymbol{H}_{\text{EH}}^{\text{los}} + \sqrt{\frac{1}{\kappa_{\text{EH}} + 1}} \boldsymbol{H}_{\text{EH}}^w\), where \(\kappa_{\text{ID}} = \kappa_{\text{EH}} = 20\) are the Rician factors, \(\boldsymbol{H}_{\text{ID}}^w\) and \(\boldsymbol{H}_{\text{EH}}^w\) are generated as CSCG random matrices with each element being zero mean and variance of \(\alpha_{\text{ID}}^2\) and \(\alpha_{\text{EH}}^2\), respectively, and \(\boldsymbol{H}_{\text{ID}}^{\text{los}} = \alpha_{\text{ID}} \boldsymbol{a}_{\text{ID}}(\theta_{\text{ID}}) \boldsymbol{a}_t^T(\theta_{\text{ID}})\) and \(\boldsymbol{H}_{\text{EH}}^{\text{los}} = \alpha_{\text{EH}} \boldsymbol{a}_{\text{EH}}(\theta_{\text{EH}}) \boldsymbol{a}_t^T(\theta_{\text{EH}})\) with \(\theta_{\text{ID}} = \frac{\pi}{6}\) and \(\theta_{\text{EH}} = \frac{2 \pi}{3}\). It is observed in Fig. 7(a) that when \(M = 6\), the R-max vertex, the E-max vertex, and the R-E tradeoff edge are not achievable. This is due to the fact that \(\mathrm{CRB}_{\text{ID},2} \to \infty\) and \(\mathrm{CRB}_{\text{EH},2} \to \infty\) when \(M > N_{\text{ID}} > 1\) (cf. Section III-A). When the required CRB $\Gamma_{\text{S}}$ becomes sufficiently large, the Pareto boundary contracts and approaches a boundary that is parallel to the CRB-axis. This is because the CRB constraint in (P2) can be easily satisfied, as if is not considered. Thus, the required CRB will hardly affect the C-R tradeoff. Next, it is observed in Fig. 7(b) that when \(M = N_{\text{ID}} = 2\), only the R-max vertex is achieved and the E-max vertex is not achievable. This is due to the fact that \(\mathrm{CRB}_{\text{ID},2}\) is finite as \(\boldsymbol{S}_{\text{ID}}\) is of full rank, but \(\mathrm{CRB}_{\text{EH},2} \to \infty\) as \(M > 1\). 

Next, we evaluate the performance of our proposed designs as compared to the following benchmark schemes based on the time division and the EMT.
\begin{itemize}
	\item \textbf{Time division protocol:} The transmission duration is divided into three orthogonal portions, \(\tau_{\text{ID}} \ge 0\), \(\tau_{\text{EH}} \ge 0\), and \(\tau_{\text{S}} \ge 0\), such that \(\tau_{\text{ID}} + \tau_{\text{EH}} + \tau_{\text{S}} = 1\), during which the H-AP employs the transmit covariance matrices \(\boldsymbol{S}_{\text{ID}}\), \(\boldsymbol{S}_{\text{EH}}\), and \(\boldsymbol{S}_{\text{S},i}\), respectively. For case \(i \in \{1,2\}\), the corresponding communication rate, harvested energy, and estimation CRB are expressed as \({R}_{\text{TD}}(\tau_{\text{ID}}, \tau_{\text{EH}}, \tau_{\text{S}}) = \tau_{\text{ID}} {R}(\boldsymbol{S}_{\text{ID}})\), \({E}_{\text{TD},i}(\tau_{\text{ID}}, \tau_{\text{EH}}, \tau_{\text{S}}) = \tau_{\text{ID}} {E}(\boldsymbol{S}_{\text{ID}}) + \tau_{\text{EH}} {E}(\boldsymbol{S}_{\text{EH}}) + \tau_{\text{S}} {E}(\boldsymbol{S}_{\text{S},i})\), and \(\mathrm{CRB}_{\text{TD},i}(\tau_{\text{ID}}, \tau_{\text{EH}}, \tau_{\text{S}}) = \mathrm{CRB}_i(\tau_{\text{ID}} \boldsymbol{S}_{\text{ID}} + \tau_{\text{EH}} \boldsymbol{S}_{\text{EH}} + \tau_{\text{S}} \boldsymbol{S}_{\text{S},i})\), respectively. The time periods \(\tau_{\text{ID}}\), \(\tau_{\text{EH}}\), and \(\tau_{\text{S}}\) are optimized to achieve different C-R-E tradeoffs.
	
	\item \textbf{EMT over the ID channel:} This scheme is motivated by the communication-optimal EMT design for MIMO communication \cite{telatar1999capacity}. For the point target case, supposing that the SVD of the ID channel \(\boldsymbol{H}_{\text{ID}}\) is \(\boldsymbol{H}_{\text{ID}} = \boldsymbol{U}_{\text{ID}} \boldsymbol{\Lambda}_{\text{ID}} \boldsymbol{V}_{\text{ID}}^H\), we design the transmit covariance matrix as \(\boldsymbol{S} = \boldsymbol{V}_{\text{ID}} \boldsymbol{\widehat{S}}^{\text{ID}} \boldsymbol{V}_{\text{ID}}^H\), where \(\boldsymbol{\widehat{S}}^{\text{ID}}\) is a diagonal matrix that is optimized based on problem (P1). For the extended target case, supposing the SVD of the projected ID channel \(\widetilde{\boldsymbol{H}}_{\text{ID}}\) in problem (P2.2) is \(\widetilde{\boldsymbol{H}}_{\text{ID}} = \widetilde{\boldsymbol{U}}_{\text{ID}} \widetilde{\boldsymbol{\Lambda}}_{\text{ID}} \widetilde{\boldsymbol{V}}_{\text{ID}}^H\), we design the transmit covariance as \(\boldsymbol{S} = \boldsymbol{V}^{\overline{\mathrm{null}}} \boldsymbol{S}_1 {\boldsymbol{V}^{\overline{\mathrm{null}}}}^H + p_0 \boldsymbol{V}^{\mathrm{null}} {\boldsymbol{V}^{\mathrm{null}}}^H = \boldsymbol{V}^{\overline{\mathrm{null}}} \widetilde{\boldsymbol{V}}_{\text{ID}} \boldsymbol{\widehat{S}}_{1}^{\text{ID}} \widetilde{\boldsymbol{V}}_{\text{ID}}^H {\boldsymbol{V}^{\overline{\mathrm{null}}}}^H + p_0 \boldsymbol{V}^{\mathrm{null}} {\boldsymbol{V}^{\mathrm{null}}}^H\) based on \eqref{S_opt2}, where \(\boldsymbol{\widehat{S}}_{1}^{\text{ID}}\) is a diagonal matrix that is optimized together with $p_0$ based on problem (P2.2).
	
	\item \textbf{EMT over combined ID and EH channels:} This scheme is motivated by the optimal solution to (P2), in which the SVD of the combined ID and EH channels \(\boldsymbol{H} = \big[\boldsymbol{H}_{\text{ID}}^T \ \boldsymbol{H}_{\text{EH}}^T\big]^T\) is employed to facilitate the optimization.  As a result, for the point target case, by using the SVD of \(\boldsymbol{H}\) in \eqref{svdH}, we design the transmit covariance matrix as \(\boldsymbol{S} = \boldsymbol{V}^{\overline{\mathrm{null}}} \boldsymbol{\widehat{S}} {\boldsymbol{V}^{\overline{\mathrm{null}}}}^H\), where \(\boldsymbol{\widehat{S}}\) is a diagonal matrix that is optimized based on problem (P1). For the extended target case, supposing the SVD of the combined projected ID and EH channels \(\widetilde{\boldsymbol{H}} = \big[\widetilde{\boldsymbol{H}}_{\text{ID}}^T \ \widetilde{\boldsymbol{H}}_{\text{EH}}^T\big]^T\) is \(\widetilde{\boldsymbol{H}} = \widetilde{\boldsymbol{U}} \widetilde{\boldsymbol{\Lambda}} \widetilde{\boldsymbol{V}}^H \), we design the transmit covariance as \(\boldsymbol{S} = \boldsymbol{V}^{\overline{\mathrm{null}}} \boldsymbol{S}_1 {\boldsymbol{V}^{\overline{\mathrm{null}}}}^H + p_0 \boldsymbol{V}^{\mathrm{null}} {\boldsymbol{V}^{\mathrm{null}}}^H = \boldsymbol{V}^{\overline{\mathrm{null}}} \widetilde{\boldsymbol{V}} \boldsymbol{\widehat{S}}_{1} \widetilde{\boldsymbol{V}}^H {\boldsymbol{V}^{\overline{\mathrm{null}}}}^H + p_0 \boldsymbol{V}^{\mathrm{null}} {\boldsymbol{V}^{\mathrm{null}}}^H\) based on \eqref{S_opt2},
	where \(\boldsymbol{\widehat{S}}_{1}\) is a diagonal matrix that is optimized together with $p_0$ based on problem (P2.2).
\end{itemize}

\begin{figure}[tb]
	\centering 
	\subfloat[\footnotesize Point target case.] {\includegraphics[width=0.45\columnwidth]{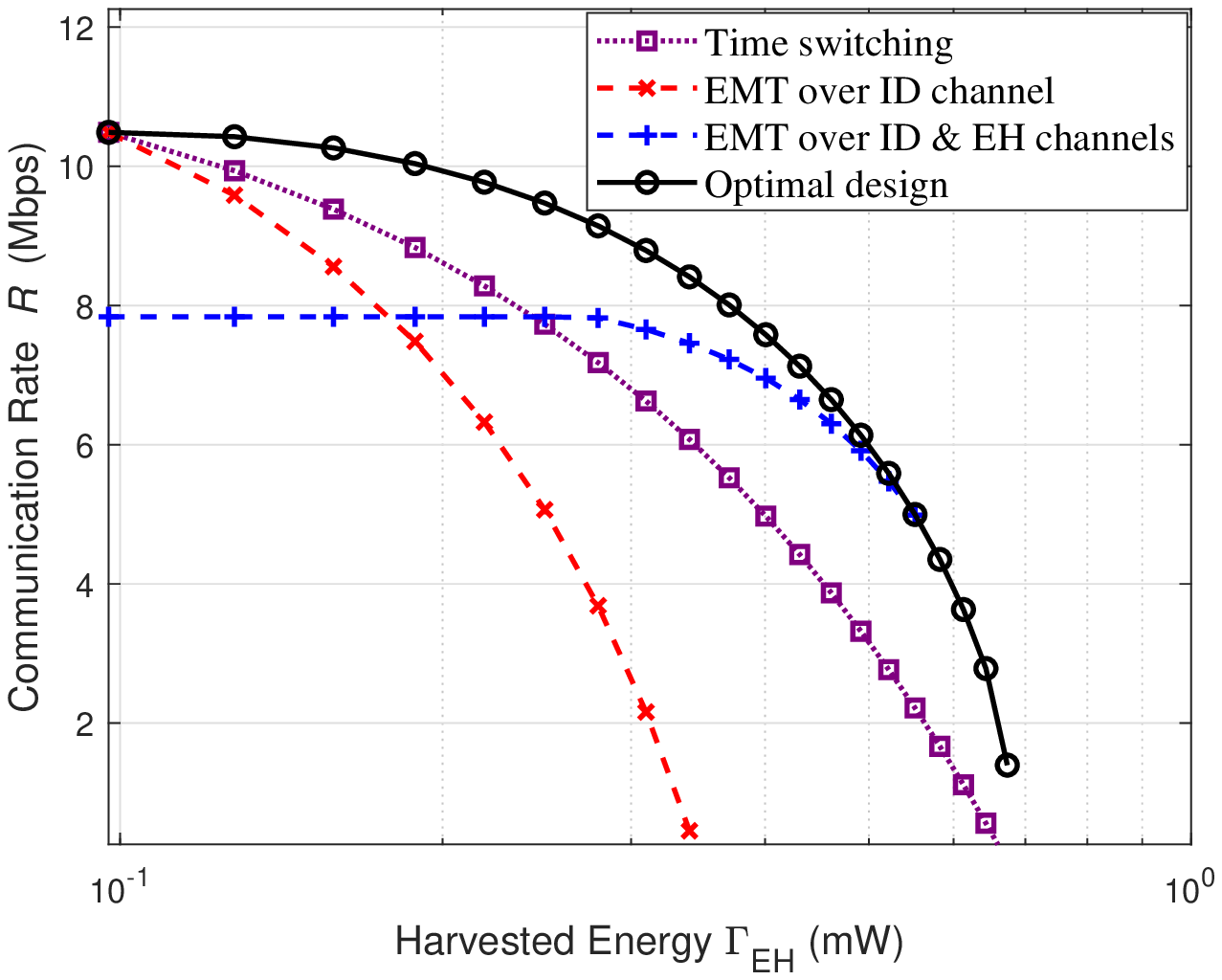}} 
	\subfloat[\footnotesize Extended target case.] {\includegraphics[width=0.45\columnwidth]{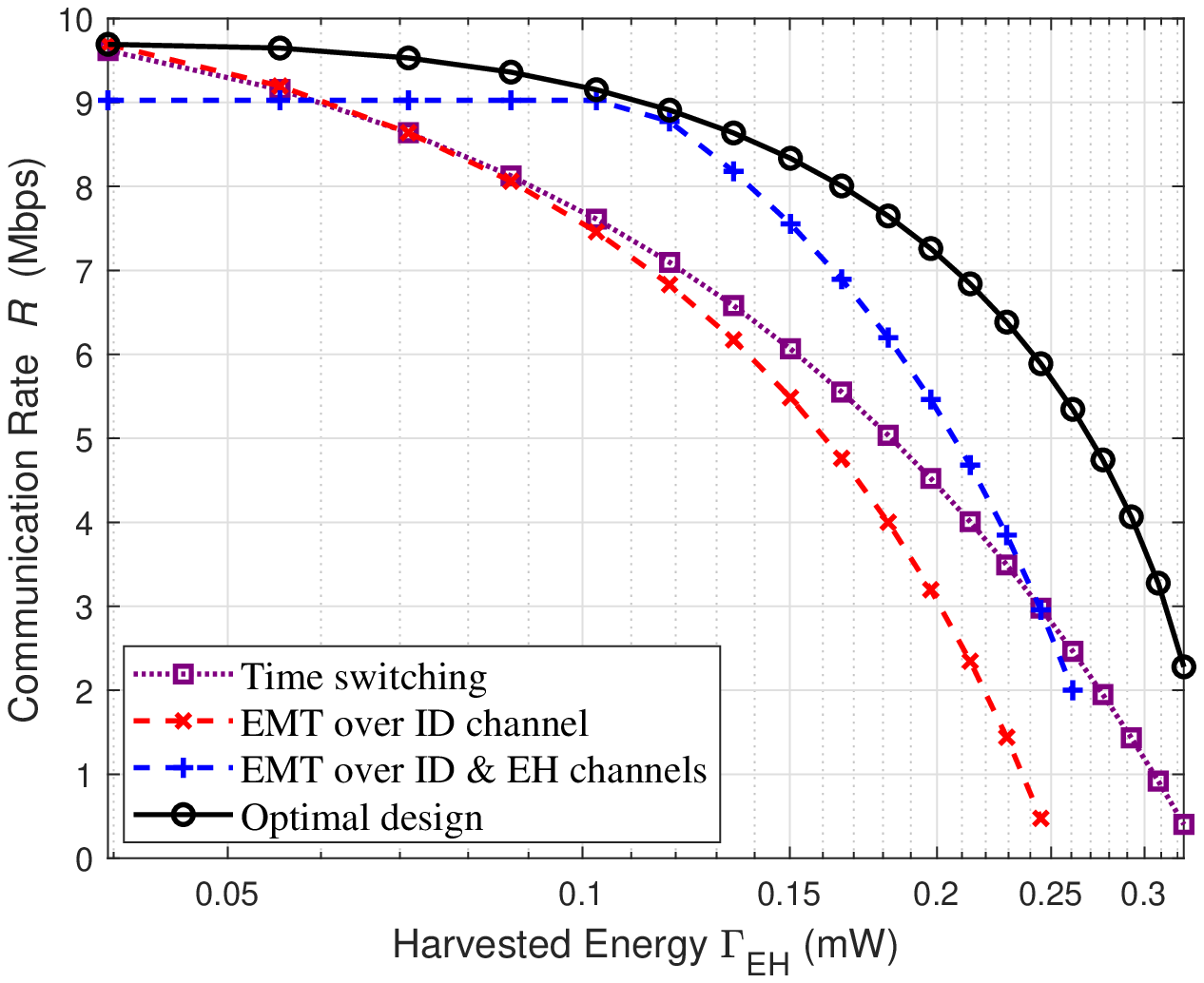}} 
	\caption{The communication rate \({R}\) versus the energy constraint \(\Gamma_{\text{EH}}\) with \(N_{\text{ID}} = N_{\text{EH}} = 2\).} 
\end{figure} 
Figs. 8(a) and 8(b) show the obtained communication rate \({R}\) by solving problems (P1) and (P2) versus the EH constraint \(\Gamma_{\text{EH}}\) for the point and extended target cases, respectively, where the sensing constraints are set as \(\Gamma_{\text{S}} = 50 \mathrm{CRB}_{\mathrm{min},1}\) and \(\Gamma_{\text{S}} = 50 \mathrm{CRB}_{\mathrm{min},2}\), respectively. Here, we set \(N_{\text{ID}} = N_{\text{EH}} = 2\) and consider Rician fading for \(\boldsymbol{H}_{\text{ID}}\) and \(\boldsymbol{H}_{\text{EH}}\). It is observed that for each target case, our proposed optimal design outperforms the other three benchmark schemes. The EMT over combined ID and EH channels is observed to outperform the other two benchmarks in the medium regime of \(\Gamma_{\text{EH}}\). This shows the benefit of considering both ID and EH channels for EMT in this case. In addition, the time division protocol and EMT over the ID channel are observed to perform close to the proposed design when \(\Gamma_{\text{EH}}\) is low. This is because the three schemes behave as the communication-optimal design in this case, as both the CRB and EH constraints become less stringent. 

\begin{figure}[tb]
	\centering 
	\subfloat[\footnotesize Point target case.] {\includegraphics[width=0.45\columnwidth]{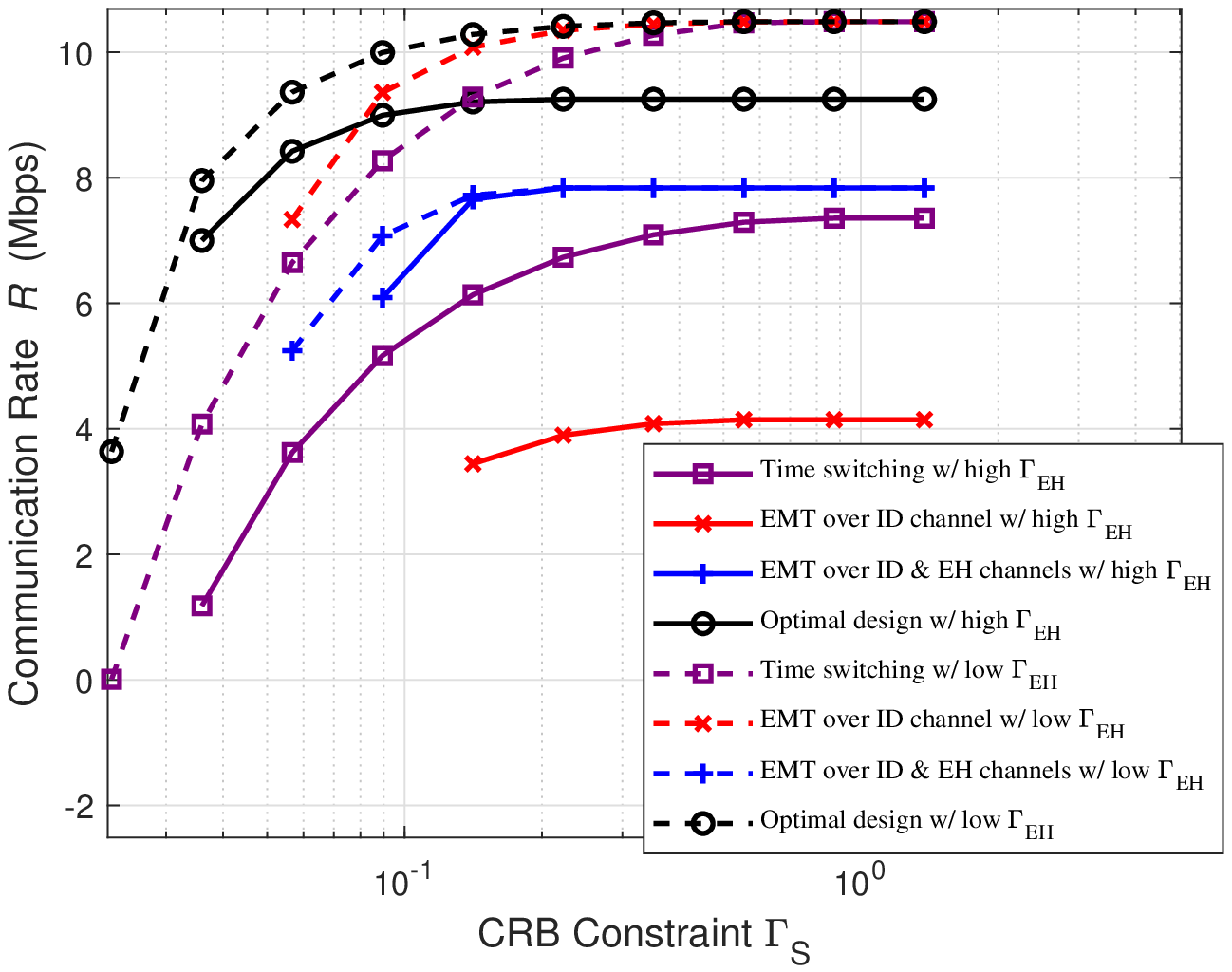}} 
	\subfloat[\footnotesize Extended target case.] {\includegraphics[width=0.45\columnwidth]{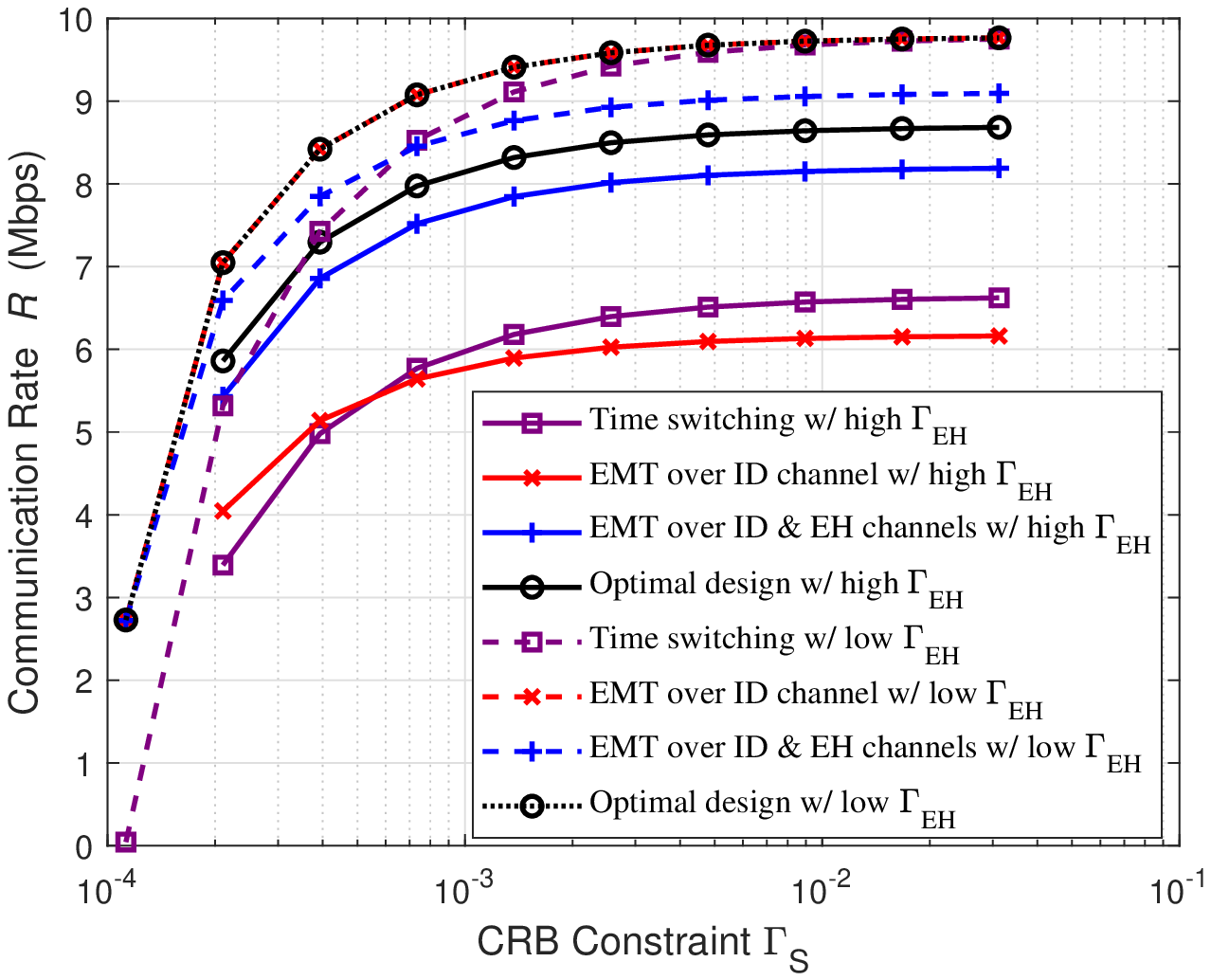}} 
	\caption{The communication rate \({R}\) versus the CRB constraint \(\Gamma_{\text{S}}\) with \(N_{\text{ID}} = N_{\text{EH}} = 2\).} 
\end{figure}
Figs. 9(a) and 9(b) show the obtained communication rate \({R}\) by problems (P1) and (P2) versus the estimation CRB constraint \(\Gamma_{\text{S}}\) for the point and extended target cases, respectively, in which two EH constraints \(\Gamma_{\text{EH}} = 0.5 {E}_{\mathrm{max}}\) (high \(\Gamma_{\text{EH}}\) case) and \(\Gamma_{\text{EH}} = 0.05 {E}_{\mathrm{max}}\) (low \(\Gamma_{\text{EH}}\) case) are considered, with \(N_{\text{ID}} = N_{\text{EH}} = 2\). It is observed that for each case with different EH constraints, our proposed optimal design outperforms the other three benchmark schemes. The EMT over combined ID and EH channels exhibits superior performance compared with EMT over the ID channel when \(\Gamma_{\text{EH}}\) is high, while the opposite results hold with low \(\Gamma_{\text{EH}}\), due to the similar reasons for the results in Figs. 8(a) and 8(b). Furthermore, with a low \(\Gamma_{\text{EH}}\), the performance of time division protocol and EMT over the ID channel are observed to approach that of the optimal design when \(\Gamma_{\text{S}}\) becomes large. This is due to the fact that both schemes can achieve the maximum communication rate with less restrictive CRB and EH constraints in this case.

\section{Conclusion}

This paper studied the fundamental C-R-E performance limits for a new multi-functional MIMO system integrating triple functions of sensing, communication, and powering. For both the point and extended target cases, we characterized the Pareto boundaries of the C-R-E performance regions, by optimally solving novel MIMO communication rate maximization problems subject to both EH and estimation CRB constraints. It was shown that for the point target case, the resultant C-R-E tradeoff highly depends on the correlations among the ID, EH, and sensing channels; while for the extended target case, the resultant C-R-E tradeoff relies on the system configuration such as the number of antennas equipped at the transceivers. Furthermore, the proposed optimal design was shown to significantly outperform the benchmark schemes based on time division and EMT. It is our hope that this paper can open new avenues for leveraging radio signals for multiple purposes towards 6G. 

\appendix

\subsection{Proof of Proposition \ref{prop1}}

Substituting \eqref{D} and \eqref{S_p}, problem \eqref{L_P_1} is re-expressed as
\begin{equation} \label{L_P_11}
	\begin{aligned}
		\max_{\boldsymbol{S}_{11},\boldsymbol{C},\boldsymbol{S}_{00}} &\ \log_2 \det \Big(\boldsymbol{I}_{N_{\text{ID}}} + \frac{1}{\sigma_{\text{ID}}^2} \boldsymbol{H}_{\text{ID}} (\boldsymbol{Q}^{\overline{\mathrm{null}}} \boldsymbol{S}_{11} {\boldsymbol{Q}^{\overline{\mathrm{null}}}}^H \\
		&\ + \boldsymbol{Q}^{\overline{\mathrm{null}}} \boldsymbol{C} {\boldsymbol{Q}^{\mathrm{null}}}^H + \boldsymbol{Q}^{\mathrm{null}} \boldsymbol{C}^H {\boldsymbol{Q}^{\overline{\mathrm{null}}}}^H \\
		&\ + \boldsymbol{Q}^{\mathrm{null}} \boldsymbol{S}_{00} {\boldsymbol{Q}^{\mathrm{null}}}^H) \boldsymbol{H}_{\text{ID}}^H\Big) - \mathrm{tr}(\boldsymbol{\Sigma}^{\overline{\mathrm{null}}} \boldsymbol{S}_{11}) \\
		\mathrm{s.t.} &\ \begin{bmatrix}
			\boldsymbol{S}_{11} & \boldsymbol{C} \\ \boldsymbol{C}^H & \boldsymbol{S}_{00}
		\end{bmatrix} \succeq \boldsymbol{0}.
	\end{aligned}
\end{equation}
According to Lemma \ref{lem1}, to solve \eqref{L_P_1}, we only need to deal with problem \eqref{L_P_11} for the case with \(\boldsymbol{H}_{\text{ID}} (\boldsymbol{Q}^{\overline{\mathrm{null}}} \boldsymbol{C} {\boldsymbol{Q}^{\mathrm{null}}}^H + \boldsymbol{Q}^{\mathrm{null}} \boldsymbol{C}^H {\boldsymbol{Q}^{\overline{\mathrm{null}}}}^H+ \boldsymbol{Q}^{\mathrm{null}} \boldsymbol{S}_{00} {\boldsymbol{Q}^{\mathrm{null}}}^H) \boldsymbol{H}_{\text{ID}}^H = \boldsymbol{0}\). Therefore, we can simply choose \(\boldsymbol{C}^* = \boldsymbol{0}\) and \(\boldsymbol{S}_{00}^* = \boldsymbol{0}\) as the optimal solution of \(\boldsymbol{C}\) and \(\boldsymbol{S}_{00}\) for obtaining the dual function \(g(\lambda, \nu, \boldsymbol{Z})\) only, and thus \eqref{L_P_11} is simplified as the optimization of \(\boldsymbol{S}_{11}\) in the following:
\begin{equation} \label{L_P_111}
	\begin{aligned}
		\max_{\boldsymbol{S}_{11} \succeq \boldsymbol{0}} &\ \log_2 \det \Big(\boldsymbol{I}_{N_{\text{ID}}} + \frac{1}{\sigma_{\text{ID}}^2} \boldsymbol{H}_{\text{ID}} \boldsymbol{Q}^{\overline{\mathrm{null}}} \boldsymbol{S}_{11} {\boldsymbol{Q}^{\overline{\mathrm{null}}}}^H \boldsymbol{H}_{\text{ID}}^H\Big) \\
		&\ - \mathrm{tr}(\boldsymbol{\Sigma}^{\overline{\mathrm{null}}} \boldsymbol{S}_{11}).
	\end{aligned}
\end{equation}
Define \(\boldsymbol{S}_{11} = (\boldsymbol{\Sigma}^{\overline{\mathrm{null}}})^{-\frac{1}{2}} \boldsymbol{V}_{\text{p}} \boldsymbol{\widetilde{S}}_{11} \boldsymbol{V}_{\text{p}}^H (\boldsymbol{\Sigma}^{\overline{\mathrm{null}}})^{-\frac{1}{2}}\). Problem \eqref{L_P_111} is re-expressed as
\begin{equation} \label{L_P_2}
	\max_{\boldsymbol{\widetilde{S}}_{11} \succeq \boldsymbol{0}} \ \log_2 \det \Big(\boldsymbol{I}_{\widetilde{r}_{\text{p}}} + \frac{1}{\sigma_{\text{ID}}^2} \boldsymbol{\Lambda}_{\text{p}}^H \boldsymbol{\Lambda}_{\text{p}} \boldsymbol{\widetilde{S}}_{11}\Big) - \mathrm{tr}(\boldsymbol{\widetilde{S}}_{11}).
\end{equation}
As shown in \cite{zhang2013mimo}, the optimal solution to problem \eqref{L_P_2} is given by \(\boldsymbol{\widetilde{S}}_{11}^* = \mathrm{diag}(\widetilde{p}_1, \dots, \widetilde{p}_{\widetilde{r}_{\text{p}}})\), where \(\widetilde{p}_k = (\frac{1}{\ln 2}-\frac{\sigma_{\text{ID}}^2}{\lambda_{\text{p},k}^2})^+\), \(\forall k \in \{1,\dots,\widetilde{r}_{\text{p}}\}\). Thus, the optimal solution of \(\boldsymbol{S}_{11}\) to \eqref{L_P_11} is
\begin{equation}
	\boldsymbol{S}_{11}^* = (\boldsymbol{\Sigma}^{\overline{\mathrm{null}}})^{-\frac{1}{2}} \boldsymbol{V}_{\text{p}} \boldsymbol{\widetilde{S}}_{11}^* \boldsymbol{V}_{\text{p}}^H (\boldsymbol{\Sigma}^{\overline{\mathrm{null}}})^{-\frac{1}{2}}.
\end{equation}
Thus, this completes the proof.

\subsection{Proof of Proposition \ref{prop3}}

To facilitate the proof of this proposition, we first present the following lemmas.

\begin{lemma} \label{lem2}
	If \(\begin{bmatrix}
		\boldsymbol{S}_1 & \boldsymbol{B} \\ \boldsymbol{B}^H & \boldsymbol{S}_0
	\end{bmatrix} \succeq \boldsymbol{0}\), then we have \(\mathrm{tr}\left(\begin{bmatrix}
		\boldsymbol{S}_1 & \boldsymbol{B} \\ \boldsymbol{B}^H & \boldsymbol{S}_0
	\end{bmatrix}^{-1}\right) \ge \mathrm{tr}\left(\begin{bmatrix}
		\boldsymbol{S}_1 & \boldsymbol{0} \\ \boldsymbol{0} & \boldsymbol{S}_0
	\end{bmatrix}^{-1}\right)\).
\end{lemma}

\textit{Proof:} Based on the Schur complement, \(\begin{bmatrix}
	\boldsymbol{S}_1 & \boldsymbol{B} \\ \boldsymbol{B}^H & \boldsymbol{S}_0
\end{bmatrix} \succeq \boldsymbol{0}\) is equivalent to \(\boldsymbol{S}_1 \succeq \boldsymbol{0}\) and \(\boldsymbol{F}_1 = \boldsymbol{S}_0 - \boldsymbol{B}^H \boldsymbol{S}_1^{-1} \boldsymbol{B} \succeq \boldsymbol{0}\), while is also equivalent to \(\boldsymbol{S}_0 \succeq \boldsymbol{0}\) and \(\boldsymbol{F}_0 = \boldsymbol{S}_1 - \boldsymbol{B} \boldsymbol{S}_0^{-1} \boldsymbol{B}^H \succeq \boldsymbol{0}\). According to the partitioned matrix formula, we have \(\begin{bmatrix}
\boldsymbol{S}_1 & \boldsymbol{B} \\ \boldsymbol{B}^H & \boldsymbol{S}_0
\end{bmatrix}^{-1} = \begin{bmatrix}
\boldsymbol{F}_0^{-1} & -\boldsymbol{S}_1^{-1} \boldsymbol{B} \boldsymbol{F}_1^{-1} \\ -\boldsymbol{F}_1^{-1} \boldsymbol{B}^H \boldsymbol{S}_1^{-1} & \boldsymbol{F}_1^{-1}
\end{bmatrix}\) and \(\begin{bmatrix}
\boldsymbol{S}_1 & \boldsymbol{0} \\ \boldsymbol{0}^H & \boldsymbol{S}_0
\end{bmatrix}^{-1} = \begin{bmatrix}
\boldsymbol{S}_1^{-1} & \boldsymbol{0} \\ \boldsymbol{0} & \boldsymbol{S}_0^{-1}
\end{bmatrix}\).
Since \(\boldsymbol{F}_1 \succeq \boldsymbol{0}\) and \(\boldsymbol{F}_0 \succeq \boldsymbol{0}\), we have \(\mathrm{tr}(\boldsymbol{F}_0^{-1}) = \mathrm{tr}(\boldsymbol{S}_1^{-1} + \boldsymbol{S}_1^{-1} \boldsymbol{B} \boldsymbol{F}_1^{-1} \boldsymbol{B}^H \boldsymbol{S}_1^{-1}) \ge \mathrm{tr}(\boldsymbol{S}_1^{-1})\) and  \(\mathrm{tr}(\boldsymbol{F}_1^{-1}) = \mathrm{tr}(\boldsymbol{S}_0^{-1} + \boldsymbol{S}_0^{-1} \boldsymbol{B}^H \boldsymbol{F}_0^{-1} \boldsymbol{B} \boldsymbol{S}_0^{-1}) \ge \mathrm{tr}(\boldsymbol{S}_0^{-1})\), respectively. Therefore, it follows that \(\mathrm{tr}\left(\begin{bmatrix}
	\boldsymbol{S}_1 & \boldsymbol{B} \\ \boldsymbol{B}^H & \boldsymbol{S}_0
\end{bmatrix}^{-1}\right) = \mathrm{tr}(\boldsymbol{F}_0^{-1}) + \mathrm{tr}(\boldsymbol{F}_1^{-1}) \ge \mathrm{tr}(\boldsymbol{S}_1^{-1}) + \mathrm{tr}(\boldsymbol{S}_0^{-1}) = \mathrm{tr}\left(\begin{bmatrix}
	\boldsymbol{S}_1 & \boldsymbol{0} \\ \boldsymbol{0} & \boldsymbol{S}_0
\end{bmatrix}^{-1}\right)\),
where the equality holds when \(\boldsymbol{B} = \boldsymbol{0}\).
\hfill \(\square\)

\begin{lemma} \label{lem3}
	Suppose that \(\boldsymbol{S}_0 \in \mathbb{S}_+^{M-r}\) with diagonal elements \(p_{0,1}, \dots, p_{0,M-r}\), where \(p_{0,k} \ge 0, \forall k \in \{1, \dots, M-r\}\). Then it follows that \(\mathrm{tr}(\boldsymbol{S}_0^{-1}) \ge \mathrm{tr}\big(\mathrm{diag}(p_{0,1}, \dots, p_{0,M-r})^{-1}\big)\).
\end{lemma}

\textit{Proof:} We decompose \(\boldsymbol{S}_0\) as \(\boldsymbol{S}_0 = \begin{bmatrix}
	\boldsymbol{S}_{0,1} & \boldsymbol{s}_{0,1} \\ \boldsymbol{s}_{0,1}^H & p_{0,M-r}
\end{bmatrix}\),
where \(\boldsymbol{S}_{0,1} \in \mathbb{S}^{M-r-1}\) and \(\boldsymbol{s}_{0,1} \in \mathbb{C}^{(M-r-1) \times 1}\). From Lemma \ref{lem2}, we have \(\mathrm{tr}\left(\begin{bmatrix}
	\boldsymbol{S}_{0,1} & \boldsymbol{s}_{0,1} \\ \boldsymbol{s}_{0,1}^H & p_{0,M-r}
\end{bmatrix}^{-1}\right) \ge \mathrm{tr}\left(\begin{bmatrix}
	\boldsymbol{S}_{0,1} & \boldsymbol{0} \\ \boldsymbol{0} & p_{0,M-r}
\end{bmatrix}^{-1}\right)\),
where \(\boldsymbol{S}_{0,1} \succeq \boldsymbol{0}\) and \(p_{0,M-r} \ge 0\). We further decompose \(\boldsymbol{S}_{0,1}\) to \(\boldsymbol{S}_{0,2} \in \mathbb{S}^{M-r-2}, \dots,\) \(\boldsymbol{S}_{0,M-r-1} = p_{0,1}\). After repeating the process above, this lemma can be proved.
\hfill \(\square\)

According to Lemma \ref{lem2}, subject to constraint in \eqref{RRRc}, the left-hand side of \eqref{RRRb} is minimized to \(\mathrm{tr}(\boldsymbol{S}_1^{-1}) + \mathrm{tr}(\boldsymbol{S}_0^{-1})\) when \(\boldsymbol{B} = \boldsymbol{0}\). We assume that \(\boldsymbol{S}_0\) is not diagonal and define \(p_{0} \triangleq \frac{\sum_{k=1}^{M-r} p_{0,k}}{M-r}\). According to Lemma \ref{lem3}, we have \(\mathrm{tr}(\boldsymbol{S}_0^{-1}) \ge \mathrm{tr}\big(\mathrm{diag}(p_{0,1}, \dots, p_{0,M-r})^{-1}\big) = \sum_{k=1}^{M-r} \frac{1}{p_{0,k}} \ge \frac{M-r}{p_{0}} = \mathrm{tr}\big((p_{0} \boldsymbol{I}_{M-r})^{-1}\big)\), where the second inequality holds based on the harmonic inequality. Thus, with \(\mathrm{tr}(\boldsymbol{S}_0)\) unchanged, \(\mathrm{tr}(\boldsymbol{S}_0^{-1})\) is minimized by removing the non-diagonal elements of \(\boldsymbol{S}_0\) and averaging its diagonal elements. Therefore, the optimal solution to problem (P2.1) satisfies that \(\boldsymbol{B} = \boldsymbol{0}\) and \(\boldsymbol{S}_0 = p_{0} \boldsymbol{I}_{M-r}\). Thus, this completes the proof.

\subsection{Proof of Proposition \ref{prop5}}

Let \(\lambda \ge 0\), \(\mu \ge 0\), and \(\nu \ge 0\) denote the dual variables associated with the constraints in \eqref{RRRRRa}, \eqref{RRRRRb}, and \eqref{RRRRRc}, respectively. The Lagrangian of problem (P3) is given by
\begin{equation}
	\begin{aligned}
		\mathcal{L}_3&(\boldsymbol{S}_1, p_{0}, \lambda, \mu, \nu)
		= - \mathrm{tr}(\boldsymbol{E} \boldsymbol{S}_1) - \mu \mathrm{tr}(\boldsymbol{S}_1^{-1}) \\
		&- (M-r) (\frac{\mu}{p_{0}} + \nu p_{0}) - \lambda \Gamma_{\text{EH}} + \mu \Gamma_{\text{S},2} + \nu P,
	\end{aligned}
\end{equation}
where \(\boldsymbol{E} \triangleq \nu \boldsymbol{I}_{r} - \widetilde{\boldsymbol{H}}_{\text{ID}}^H \widetilde{\boldsymbol{H}}_{\text{ID}} - \lambda \widetilde{\boldsymbol{H}}_{\text{EH}}^H \widetilde{\boldsymbol{H}}_{\text{EH}}\). Accordingly, the dual function of (P3) is defined as
\begin{equation} \label{dualfunc_e}
	g_3(\lambda, \mu, \nu) = \max_{\boldsymbol{S}_1 \succeq \boldsymbol{0}, p_{0} \ge 0} \ \mathcal{L}_3(\boldsymbol{S}_1, p_{0}, \lambda, \mu, \nu).
\end{equation} 
We have the following lemma, which can be verified similarly as Lemma \ref{lem1}.
\begin{lemma} \label{lem4}
	In order for the dual function \(g_3(\lambda, \mu, \nu)\) to be bounded from above, it must hold that \(\boldsymbol{E} \succeq \boldsymbol{0}\).
\end{lemma}

Based on Lemma \ref{lem4}, the dual problem of (P3) is defined as
\begin{equation}
	(\text{D}3):\min_{\lambda \ge 0, \mu \ge 0, \nu \ge 0} \ g_3(\lambda, \mu, \nu), \ \mathrm{s.t.} \ \boldsymbol{E} \succeq \boldsymbol{0}.
\end{equation}
Since (P3) is convex and satisfies the Slater’s condition, strong duality holds between (P3) and its dual problem (D3) \cite{boyd2004vandenberghe}. Therefore, we can solve (P3) by equivalently solving (D3).

First, we find the dual function \(g_3(\lambda, \mu, \nu)\) in \eqref{dualfunc_e} under given \(\lambda \ge 0\), \(\mu \ge 0\), and \(\nu \ge 0\). By dropping the constant terms \(- \lambda \Gamma_{\text{EH}} + \mu \Gamma_{\text{S},2} + \nu P\), the problem in \eqref{dualfunc_e} is equivalent to
\begin{equation} \label{L_E_1}
	\max_{\boldsymbol{S}_1 \succeq \boldsymbol{0}, p_{0} \ge 0} \ - \mathrm{tr}(\boldsymbol{E} \boldsymbol{S}_1) - \mu \mathrm{tr}(\boldsymbol{S}_1^{-1}) - (M-r) (\frac{\mu}{p_{0}} + \nu p_{0}).
\end{equation} 
The EVD of \(\boldsymbol{E}\) is expressed as \(\boldsymbol{E} = \boldsymbol{Q}_{\text{e}} \boldsymbol{\Sigma}_{\text{e}} \boldsymbol{Q}_{\text{e}}^H\), where \(\boldsymbol{\Sigma}_{\text{e}} = \mathrm{diag}(\sigma_{\text{e},1}, \dots, \sigma_{\text{e},r})\). Without loss of generality, we assume \(\sigma_{\text{e},1} \ge \dots \ge \sigma_{\text{e},r} \ge 0\). Define \(\boldsymbol{S}_1 = \boldsymbol{Q}_{\text{e}} \boldsymbol{\widetilde{S}}_{1} \boldsymbol{Q}_{\text{e}}^H\). Problem \eqref{L_E_1} is re-expressed as
\begin{equation} \label{L_E_2}
	\max_{\boldsymbol{\widetilde{S}}_{1} \succeq \boldsymbol{0}, p_{0} \ge 0} \ - \mathrm{tr}(\boldsymbol{\Sigma}_{\text{e}} \boldsymbol{\widetilde{S}}_{1}) - \mu \mathrm{tr}(\boldsymbol{\widetilde{S}}_{1}^{-1}) - (M-r) (\frac{\mu}{p_{0}} + \nu p_{0}).
\end{equation}
According to Lemma \ref{lem3} in Appendix B, the optimal solution of \(\boldsymbol{\widetilde{S}}_{1}\) to problem \eqref{L_E_2} is diagonal, i.e., we have \(\boldsymbol{\widetilde{S}}_{1} = \mathrm{diag}(\widetilde{p}_{1,1}, \dots, \widetilde{p}_{1,r})\). Therefore, \eqref{L_E_2} is re-expressed as
\begin{equation} \label{L_E_3}
	\max_{\{\widetilde{p}_{1,k} \ge 0\}, p_{0} \ge 0} \ - \sum_{k=1}^{r} (\sigma_{\text{e},k} \widetilde{p}_{1,k} + \frac{\mu}{\widetilde{p}_{1,k}}) - (M-r) (\frac{\mu}{p_{0}} + \nu p_{0}).
\end{equation}

Notice that for the case with \(\sigma_{\text{e},r} = 0\), i.e., \(\boldsymbol{E}\) is rank-deficient, the optimality of problem \eqref{L_E_3} cannot be achieved unless \(\mu = 0\). However, it is easy to verify that at the optimality of problem (P3), the constraints \eqref{RRRRRb} and \eqref{RRRRRc} should be tight. Therefore, it follows that \(\mu > 0\) and \(\nu > 0\), which means that we should have \(\sigma_{\text{e},r} > 0\), i.e., the dual problem (D3) should also satisfy the constraint \(\boldsymbol{E} \succ \boldsymbol{0}\) to achieve the optimality.

By checking the first derivative, the optimal solution to problem \eqref{L_E_3} is
\begin{align}
	\widetilde{p}_{1,k}^* &= \sqrt{\frac{\mu}{\sigma_{\text{e},k}}}, \forall k \in \{1, \dots, r\}, \label{p1*}\\
	p_{0}^* &= \sqrt{\frac{\mu}{\nu}}. \label{p0*}
\end{align}
Thus, the optimal solution of \(\boldsymbol{S}_1\) to problem \eqref{L_E_1} is
\begin{equation} \label{S1*}
	\boldsymbol{S}_1^* = \boldsymbol{Q}_{\text{e}} \mathrm{diag}(\widetilde{p}_{1,1}^*, \dots, \widetilde{p}_{1,r}^*) \boldsymbol{Q}_{\text{e}}^H.
\end{equation}

Next, we solve the dual problem (D3), which is convex but not necessarily differentiable, and thus can be solved by applying the subgradient-based method such as the ellipsoid method. For the objective function \(g_3(\lambda, \mu, \nu)\),  the subgradient at \((\lambda, \mu, \nu)\) is \(\Big[\mathrm{tr}(\widetilde{\boldsymbol{H}}_{\text{EH}}^H \widetilde{\boldsymbol{H}}_{\text{EH}} \boldsymbol{S}_1^*) - \Gamma_{\text{EH}},
\Gamma_{\text{S},2} - \mathrm{tr}\big((\boldsymbol{S}_1^*)^{-1}\big) - (M-r) \frac{1}{p_{0}^*},
P - \mathrm{tr}(\boldsymbol{S}_1^*) - (M-r) p_{0}^*\Big]^T\).
Furthermore, let \(\boldsymbol{q}_3\) denote the eigenvector of \(\boldsymbol{E}\) corresponding to its minimum eigenvalue \(\sigma_{\text{e},r}\). With the similar analysis as in Section IV, the subgradient of constraint  \(\boldsymbol{E} \succ \boldsymbol{0}\) at \((\lambda, \mu, \nu)\) is \(\big[\boldsymbol{q}_3^H \widetilde{\boldsymbol{H}}_{\text{EH}}^H \widetilde{\boldsymbol{H}}_{\text{EH}} \boldsymbol{q}_3, 0, -1\big]^T\).

Finally, we present the optimal solution to the primal problem (P3). With the optimal dual variables \(\lambda^*\), \(\mu^*\), and \(\nu^*\) at hand, the corresponding optimal solutions \(\boldsymbol{S}_1^*\) and \(p_{0}^*\) in \eqref{S1*} and \eqref{p0*} to problem \eqref{L_E_1} can be directly used for constructing the optimal primal solutions to (P3), denoted by \(\boldsymbol{S}_1^{\mathrm{low}}\) and \(p_{0}^{\mathrm{low}}\), respectively. This thus completes the proof of this proposition.

\bibliographystyle{IEEEtran}
\bibliography{SWIPTISAC}
	
\end{document}